\documentclass[a4paper,fleqn,usenatbib]{mnras}

\usepackage{newtxtext,newtxmath}

\usepackage[T1]{fontenc}
\usepackage{ae,aecompl}

\usepackage{graphicx}	
\usepackage{amsmath}	
\usepackage{amssymb}	

\title[Nova Distances With Gaia DR2]{The Distances to Novae As Seen By Gaia}

\author[B. E. Schaefer]{
Bradley E. Schaefer$^{1}$\thanks{E-mail: schaefer@lsu.edu}
\\
$^{1}$Department of Physics and Astronomy, Louisiana State University, Baton Rouge, Louisiana, 70820, USA\\
}

\date{Accepted XXX. Received YYY; in original form ZZZ}

\pubyear{2018}

\begin{document}
\label{firstpage}
\pagerange{\pageref{firstpage}--\pageref{lastpage}}
\maketitle

\begin{abstract}

The {\it Gaia} spacecraft has just released a large set of parallaxes, including 41 novae for which the fractional error is $<$30\%.  I have used these to evaluate the accuracy and bias of the many prior methods for getting nova-distances.  The best of the prior methods is the geometrical parallaxes from {\it HST} for just four novae, although the real error bars are 3$\times$ larger than stated.  The canonical method for prior nova-distances has been the expansion parallaxes from the nova shells, but this method is found to have real 1-sigma uncertainty of 0.95 mag in the distance modulus, and the prior quoted error bars are on average 3.6$\times$ worse than advertised.  The many variations on the `maximum-magnitude-rate-of-decline' (MMRD) relation are all found to be poor, too poor to be useable, and even to be non-applicable for 5-out-of-7 samples of nova, so the MMRD should no longer be used.  The prior method of using various measures of the extinction from the interstellar medium have been notoriously bad, but now a new version by \"{O}zd\"{o}nmez and coworkers has improved this to an unbiased method with 1-sigma uncertainty of 1.14 mag in the distance modulus.  For the future, I recommend in order (1) using the {\it Gaia} parallax, (2) using the catalog of \"{O}zd\"{o}nmez, (3) using $M_{max}$ = -7.0$\pm$1.4 mag as an empirical method of poor accuracy, and (4) if none of these methods is available, then to not use the nova for purposes where a distance is needed.
 
\end{abstract}




\section{Introduction}

A ubiquitous and deep problem of high importance throughout the last century of astrophysics has been measuring the distances to objects.  Realistic distances are critical to understanding the structure and organization of the objects, while the inverse-square dependency of the luminosities and energies on the distances means that any physical model must have good distances.  For novae, the last century has also featured much effort and debate to get distances.  

Many methods have been used to estimate nova distances.  With geometrical parallaxes ($\varpi$)
 not possible (except for four of the nearest novae as viewed with the {\it Hubble Space Telescope}, {\it HST}), the standard has been to use expansion parallaxes for the few novae with shells.  But even this standard was known to be poor for multiple reasons, and it could not be applied to most novae.  For most novae, with no other possibilities, workers could only make order-of-magnitude distance estimates based on various measures of the interstellar extinction, with such methods being notoriously poor.  (But \"{O}zd\"{o}nmez and coworkers have recently found ways to make this work, see below.)  With these poor calibrations, workers attempted to find a relation between the nova's absolute magnitude at peak and the speed of decline (called the `maximum-magnitude-rate-of-decline' relation, or the 'MMRD').  But this correlation has a huge scatter making the method largely useless, while recently the very existence of the relation has been disproven for novae in M31 and M87.  Another method from the past few years has been to get blackbody distances to the secondary stars.  But this method is untested, and is applicable to only a half-dozen or so systems.  So in all, until now, the all-important distances to novae are poorly known.

Now, with the public release of many accurate parallaxes from the {\it Gaia} spacecraft, we finally have confident and accurate distances to many novae.  Suddenly, we can evaluate all the prior nova distances and their methods.  This has vital relevance for future nova studies because {\it Gaia} will only get good parallaxes for less than 20\% of the known novae.  For the other 80\%, we still have to rely on the many other methods, so it is good to learn the real accuracy and biases of each method.  Thus, a primary purpose of this paper is to evaluate the prior published nova distances as based on the {\it Gaia} `ground-truth'.  

\section{Observations}

The European Space Agency {\it Gaia} satellite is awesome in its capabilities for astrometry, for getting parallax and proper motions of stars far out into our Milky Way galaxy.  The second data release (DR2) has just come out, with 22 months of operational data covering 1693 million stars from magnitude 3 to 21 (Lindegren et al. 2018).  DR2 is publicly available on-line{\footnote{https://archives.esac.esa.int/gaia}}.  DR2 includes positions, proper motions, and parallaxes (five astrometric parameters), with all sources treated as single stars.  No binary motion or source confusion is allowed for, with these being covered in later data releases.  The distribution of errors in the parallax is essentially a perfect Gaussian with the quoted 1-sigma error bars (Luri et al. 2018).  The standard uncertainty for $\varpi$ is 0.041 milli-arc-seconds (mas) for 12 mag stars, 0.057 mas for 16 mag stars, and 0.651 mas for 20 mag stars.

The traditional equation to derive the distance, $D$ in parsecs, from the observed parallax, $\varpi$ in mas, is $D=1000/\varpi$.  But various workers have long realized that the conversion from parallax to distance is really much more complex and subtle.  The trouble comes when the observed parallax is small when compared to its uncertainty.  At its extreme, a zero parallax would translate to an infinite $D$, with this being unphysical, while a perfectly good {\it negative} parallax is meaningless.  A common problem is that the uncertainty in distance will have a substantially non-Gaussian shape, with the distribution being greatly skewed when the uncertainty in the parallax, $\sigma_{\varpi}$, grows to a substantial fraction of the parallax.  For example, with an observed parallax of 0.5$\pm$0.5 mas, the 1-sigma range in parallax is 0.0 to 1.0 mas, yet the same 1-sigma range in distance is from 1000 pc out to infinity.  And the simple equation does not tell us how to handle the negative part of the parallax's distribution.  These problems become non-trivial for cases where $\sigma_{\varpi}$/$\varpi$$\gtrsim$20\% or so (Bailer-Jones 2015).

The solution to the inversion problem (i.e., to go from a measured parallax to the best distance with realistic error bars) is now know to require some appropriate assumption about the distance distribution, known as a `prior', within a Bayesian analysis.  An excellent explanation and tutorial is presented in Bailer-Jones (2015).  The solution is to adopt a prior where the {\it a priori} probability volume density decreases asymptotically to zero at infinity.  A reasonable function for the prior is an exponential decline with some appropriate distance scale.  The official {\it Gaia} DR2 publication (Luri et al. 2018) explicitly endorses this `exponentially decreasing space density' (EDSD).  With this, the probability distribution of $D$ is given by equation 18 of Bailer-Jones (2015), and I have performed the integrals on this unnormalized posterior to define the 1-sigma intervals containing the central 68.3\% probability.  The best estimate distance is given by equation 19 of Bailer-Jones.  When $\sigma_{\varpi}$/$\varpi$ rises above 0.30, two modes appear in the posterior, one of which is `data-dominated' and the longer distance is `prior-dominated', so by the time the fractional error rises above 0.373 there is a sudden increase in the mode.

So, given the {\it Gaia} values for $\varpi$ and $\sigma_{\varpi}$ along with the EDSD, the only question is the appropriate length scale.  Here I have taken the length scale to be $150/\sin({l})$ parsecs, with a maximum of 8000 pc, where ${l}$ is the galactic latitude, for a disk population.  For a halo population, I adopt a length scale of 8000 pc.  

Let us see how all this works for some schematic cases:  For a case with a length scale of 1000 pc, a measured parallax of 10.0$\pm$0.1 mas gives a $D$ with 1-sigma error bars of 100$^{+1}_{-1}$ pc, 1.0$\pm$0.1 mas gives 1010$^{+142}_{-76}$ pc, 0.1$\pm$0.1 mas gives 4600$^{+2300}_{-730}$ pc, and -0.1$\pm$0.1 mas gives 6200$^{+2700}_{-1050}$ pc.  For a case with a length scale of 8000 pc, 10.0$\pm$0.1 mas gives 100$^{+1}_{-1}$ pc, 1.0$\pm$0.1 mas gives 1019$^{+147}_{-77}$ pc, 0.1$\pm$0.1 mas gives 14300$^{+21100}_{-3300}$ pc, and -0.1$\pm$0.1 mas gives 21500$^{+21000}_{-5300}$ pc.

The {\it Gaia} Data Release 1 (DR1) has already returned the parallaxes for three novae (V603 Aql, RR Pic, and HR Del), as based on comparisons of {\it Tycho} positions plus early-epoch {\it Gaia} positions (Ramsay et al. 2017).  But these early DR1 results have quoted error bars over a full order-of-magnitude larger than the DR2 results.  Nevertheless, the Ramsay et al. study provided the first look at nova-distances with {\it Gaia}, showing that the short distance scale to SS Cyg was correct, and providing the first indications that the prior {\it HST} parallaxes and the expansion parallaxes were greatly worse than advertised.

I have examined 120 novae for reliable inclusion in the {\it Gaia} DR2 data base.  A total of 64 novae are included in Table 1.  These are divided into three samples, which I label as the `Gold', `Silver', and `Bronze' samples.  The 26 Gold novae are those with very well observed light curves from the SSH sample (Strope, Schaefer, \& Henden 2010; SSH) for which {\it Gaia} has a confident detection and a parallax with less than 30\% error.  The SSH sample of novae contains the 93 all-time best observed nova light curves, all with exhaustive light curve information collected together and systematically analyzed for the various needed light curve properties.  The Gold sample is the best and most reliable, mainly because these novae are generally the nearest and brightest.  The 15 Silver novae are mostly well observed events which are not included in SSH for various reasons, and for which {\it Gaia} returns a confident identification with a parallax error of $<$30\%.  The 41 novae in the Gold+Silver sample comprises all the confident and accurate novae parallaxes, and this is my basic group for testing the prior distances.   The 23 Bronze novae are those for which there is a reliable {\it Gaia} detection, but for which the quoted parallax error bar is $>$30\%.  These novae have no real utility for testing prior distance measures.  However, there is information in the {\it Gaia} parallax measures, but only for statistical purposes.

Many novae are not included into Table 1, for many reasons.  Three recurrent novae with red giant companion stars (T CrB, RS Oph, and V3890 Sgr) do not have reliable {\it Gaia} parallaxes because their long-period binary orbits will cause the center-of-light to wobble with shifts comparable to the parallaxes, so we must await a full solution with a later data release.  The unique nova V445 Pup is recognized in the DR2 catalog, but no parallax is recorded.  Eleven old novae (including DO Aql, V5592 Sgr, and V1213 Cen) have more than one candidate around the correct position, but I cannot decide with any useable confidence as to which (if any) DR2 objects are the real quiescent counterparts.  Twenty-six old novae (including V2274, V2362, and V2467 Cyg, plus V2264, V2295, V2313, and V2540 Oph) have no confidently identified counterpart (often likely because the counterpart is very faint) so no DR2 object can be taken as a reliable counterpart.  Fifteen old novae (including Nova Sco 1437, V728 Sco, V977 Sco, and V1187 Sco) have counterparts that are not seen by {\it Gaia}.

For possible inclusion in Table 1, I have looked at nearly all the known galactic novae for which even poor light curves are available and for which a counterpart is known.  For the galactic novae with a confident counterpart in {\it Gaia} DR2 with $\sigma_{\varpi}/\varpi<30\%$, that is the Gold and Silver samples, I think that Table 1 is complete.  For the galactic novae with a confident counterpart with $\sigma_{\varpi}/\varpi>30\%$, that is the Bronze sample, I think that Table 1 is nearly complete, while possibly missing some obscure novae.

I have included in the Silver sample two unexpected nova, both being well-known cataclysmic variable (CV) systems with dwarf nova (DN) eruptions.  Both Z Cam and AT Cnc were discovered to have expanding nova shells pointing with confidence to classical nova eruptions within previous centuries, plus the historical identification of `guest stars' in ancient chronicles (Shara et al. 2007; 2012a; 2016).  There is little light curve information for these old novae.  Still, they are useful because they represent two more cases for the small set of systems with observed classical nova (CN) eruptions as well as DN, and these nova systems now have reliable distances and absolute magnitudes in quiescence, with application to testing the `hibernation model' (Shara et al. 1986) of CV evolution.

Table 1 lists all the novae in the Gold, Silver, and Bronze samples, plus many properties for each nova.  My primary reference is my nova light curve catalog (Strope, Schaefer, \& Henden 2010, SSH), as this contains a comprehensive and uniform measure of all light curve information for the 93 best-observed novae of all time.  This is exactly what is needed for many of the tests of prior distance measures.  Other primary reference sources are Schaefer (2010) for recurrent novae (RN), Schaefer \& Patterson (1983) for BT Mon, Schaefer et al. (2013) for T Pyx, Salazar et al. (2017) for V1017 Sgr, Schaefer \& Collazzi (2010) for the V1500 Cyg class of novae, and Pagnotta \& Schaefer (2014) for many light curves and properties.  Further primary reference sources as compilations of many nova properties include the three wonderful and comprehensive papers of Shafter (1997), Duerbeck (1981), and \"{O}zd\"{o}nmez et al. (2018), plus the on-line CV Catalog of Downes, Webbink, \& Shara (1997).  For light curve information, for example for the recent nova V392 Per, I have made extensive use of the light curves of the {\it American Association of Variable Star Observers} (AAVSO).  For particular novae where there is some gap in the information, or where the sources put forth conflicting values, I have extensively consulted the original papers with the observations.

In Table 1, column 1 lists the nova.  Column 2 lists the sample, either Gold, Silver, or Bronze.  Column 3 gives the nova type, with the basic division being the `classical novae' (CN) and the `recurrent novae' (RN).  Various additional divisions are included, for example the notation `DN' indicates that the nova system has been seen to experience dwarf nova eruptions.  (To anticipate, these novae are indistinguishable from the other novae in terms of their absolute magnitude in quiescence, with this violating a prediction of the `hibernation model' for the evolution of CVs.)  `Hi-$\Delta$m' notates that the star is a V1500 Cyg system, where the long-post-eruption quiescence magnitude is over 2.5 mag brighter than the pre-eruption magnitude (Schaefer \& Collazzi 2010).  (To anticipate, what I will find is that the V1500 Cyg systems are greatly less luminous in quiescence than all other novae systems.)  I also notate for V838 Her that it has been identified by Pagnotta \& Schaefer (2014) as being a likely RN that has had multiple eruptions in the last century but with only one such discovered.  Further, I note that AR Cir {\it might} be a symbiotic system (i.e., have a red giant companion star), and {\it might} have had a symbiotic nova eruption.  Column 5 completes the block describing the novae by giving the light curve classification from SSH.  `S' denotes novae with a smooth light curve, `P' is for light curves with distinct plateau around the transition phase, `O' class novae show quasi-periodic oscillations around the transition phase, `D' novae are those with a deep dust dip in the light curve, `F' novae display a long flat top at the maximum of their light curve, `J' novae have large flares or jitters in their light curve around the time of maximum, and `C' novae have a distinct cusp with a slow and accelerating rise to a second maximum followed by a fast fall.

The next block of Table 1, columns 6 and 7, give the new {\it Gaia} input.  This is the measured parallax and its 1-sigma error bar (in  units of milli-arcseconds), and the derived distance (in units of parsecs).  Again, the distances are derived with the EDSD Bayesian prior, and the quoted error bars display the central 68.3\% of the probability distribution.

The next block of Table 1 contains 6 columns with light curve information.  Column 8 reports the peak magnitude, $V_{max}$.  For two systems (Z Cam and AT Cnc) for which the nova is only known from ancient historical records (as well as from their expanding shells), the peak is not known, but it must have been something like 0$\pm$3 mag.  It is difficult to define a formal error bar even for the well-observed light curves from SSH, but a typical real uncertainty is roughly 0.1--0.2 mag.  Some of the light curves in the Silver and Bronze samples are poorly sampled and the real error bars for $V_{max}$ can be more like 0.2--0.5 mag.  For V1017 Sgr and CT Ser, the peaks were apparently missed, so there is large uncertainty in $V_{max}$, as indicated in column 8.  Column 9 gives the magnitude at a time 15 days after peak, $V_{15d}$, and the uncertainties are only larger than for $V_{max}$.  An additional problem is knowing the date of the peak (especially in the case of the J-class novae), and often the light curve 15-days after the maximum is fast fading, so large changes in $V_{15d}$ result from modest uncertainties in the peak date.  Column 10 gives the average quiescent magnitude, $V_q$.  I have taken pre-eruption magnitudes (Collazzi et al. 2009) in preference when available.  (This is important for the V1500 Cyg stars, as the absolute magnitude before eruption is better to show the accretion rate appropriate for the long-term evolution.)  The ubiquitous magnitude variations in quiescence are roughly 0.5--2.0 mag, so with the inevitable poor sampling, it will be hard even to define the average with much accuracy.  The uncertainties in $V_q$ are hard to know, although typical error bars might be $\pm$1 mag.  Column 11 is the V-band extinction ($A_V$) from the interstellar medium.  Most of the tabulated extinctions are from compilations involving multiple measures from a wide variety of methods.  I have converted reported $E(B-V)$ values to $A_V$ as $3.1\times E(B-V)$, as appropriate for the local dust in our Milky Way's disk.  Again, formal error bars for extinctions are hard to get, with typical uncertainties being perhaps 0.1--0.3 or perhaps 10\%--30\%.  Columns 12 and 13 list the values for $t_2$ and $t_3$, given in units of days, defined as the time it takes the light curves to fall by 2.0 or 3.0, respectively, mag from the peak.  Again, formal error bars are difficult, even for well-sampled S-class light curves, and real error bars might be 10\%--30\%.  For novae with substantial jitters or with poorly sampled light curves, the real error bars can easily be 30\%--50\%.

The last block of Table 1 is the derived absolute magnitudes at maximum ($M_{max}$) and in quiescence ($M_q$).  These are calculated from the tabulated values with the absolute magnitude equalling $V-A_V-5\log[D]+5$.

\section{Testing Prior Distances To Novae}

For comparing the prior nova-distances with those from {\it Gaia}, we need quantitative measures of the errors in the distance.  The size the distance errors can be quantified as some function of either $D_{prior}-D_{Gaia} = \Delta D$ or $D_{prior}/D_{Gaia}$, where $D_{prior}$ is the pre-{\it Gaia} distance from a set being tested in this paper and $D_{Gaia}$ is the distance from the {\it Gaia} parallaxes.  The errors in these will be dominated by the prior measures, and these are often with asymmetrical distributions, such that the quantity $\log_{10}[D_{prior}/D_{Gaia}]$ will usually have a more symmetric distribution.  With a common use for the nova distances being to get luminosities and energetics, a useful measure of the distance error will be the error in the distance modulus, $\Delta \mu$.  We have $\Delta \mu = 5\times \log_{10}[D_{prior}/D_{Gaia}]$.  This quantity tells us the error in the absolute magnitude arising due to the error in the prior distance to the nova.  This can be related to the fractional distance error as $F=\Delta D / D_{Gaia} = 10^{\Delta \mu / 5}-1$.  The $F$ value loses its simple meaning as a symmetric measure when $F$ starts getting large, so the $\Delta \mu$ statistic is the general solution. 

With multiple independent measures of $\Delta \mu$ for a set of prior nova distances, the RMS scatter will equal the average measurement error for the set.  Note, this RMS scatter will be about some best fit average, so that if the prior distances have a substantial bias, the real errors in the prior distances will be substantially larger.  In general, the bulk of this scatter in $\Delta \mu$ comes from the uncertainty in the prior distance, so we can adopt the RMS scatter, $\sigma_{\Delta \mu}$ as the easy-to-calculate measure of the 1-sigma error bar for the average of the prior distances.  The prior distances might be biased, either long or short.  This can be quantified by the average differences in the distance moduli, $\langle \Delta \mu \rangle$.

We also need a measure of the size of the error bars for the prior distances.  I will use the measure $\chi \equiv \Delta D / \sigma$, with a similarity of meaning as in the usual chi-square summation.  The total error bar in the difference $\Delta D$ is $\sigma^2 = \sigma^2_{D_{prior}} + \sigma^2_{D_{Gaia}}$.  If the quoted error bars are accurate, then the distribution of the observed $\chi$ values for a set of nova distances should have an RMS scatter of near unity.  If the error bars are systematically smaller than the real scatter in the distance errors, then the RMS will be much larger than unity.  So $\sigma _{\chi}$ is a measure of the relative size of the quoted error bars with respect to the real error bars.

Of the various possible measures of the errors, individual novae can be quantified with $\Delta \mu$, while the average real errors of the collection of distances can be represented by $\sigma_{\Delta \mu}$, the size of the reported error bars can be expressed by $\sigma_{\chi}$, and the bias in the reported distances is $\langle \Delta \mu \rangle$.  An unbiased set of prior distances with $\sigma_F$=10\% errors will have $\sigma_{\Delta \mu}$ near 0.21 mag, $\sigma_{\chi}$$\cong$1.0, and $\langle \Delta \mu \rangle$$\cong$0.0.

For some applications, it is useful to define similar statistics involving the parallax itself (as opposed to the distance), with this enjoying the advantage that the error bars in parallax have a good Gaussian distribution.  The statistic $\Delta \mu = 5\times \log_{10}[\varpi_{Gaia}/\varpi_{prior}]$ can be easily calculated.  With this, we can get $\langle \Delta \mu \rangle$ and $\sigma_{\Delta \mu}$.  We can also define the fractional error in parallax as $\mathcal{F}=(\varpi_{prior}-\varpi_{Gaia})/\varpi_{Gaia}$, with this being similar to $F$.  As a measure of the size of the error bars, we can use the statistic $\psi \equiv (\varpi_{prior}-\varpi_{Gaia})/\sigma_{\varpi}$, with $\sigma_{\varpi}^2=\sigma^2_{\varpi_{prior}} + \sigma^2_{\varpi_{Gaia}}$.  The RMS value of $\psi$ for a set of novae parallaxes will be $\sigma_{\psi}$, with this being similar and close to $\sigma_{\chi}$.  So the quality of a set of prior novae parallaxes can be evaluated with the three values for $\langle \Delta \mu \rangle$ (a measure of bias high-or-low), $\sigma_{\Delta \mu}$ (measure of the real 1-sigma error in the associated magnitude), and $\sigma_{\psi}$ (a measure of the size of the reported error bars).

In this section, I will systematically check prior nova distances from many different methods.  I will not address the important sets of novae that reside in other galaxies (LMC, M31, and M87) for which many workers have derived distances independent of the novae.

\subsection{Testing Prior Parallaxes}

For the past history of astronomy, the geometrical parallax has always been the `gold standard'.  Unfortunately, few novae can get any useable parallax, at least before {\it Gaia}.  The only prior useable parallaxes were measured with the {\it Fine Guidance Sensors}, FGS, on the {\it HST} (Benedict, McArthur, Nelan, \& Harrison 2016).  Harrison et al. (2013) report on FGS parallaxes for four of the brightest and nearest classical novae.  

The FGS measured parallaxes are given in Table 2 and Figure1, along with the measured parallaxes of {\it Gaia} DR2.  I will make the comparison between the parallaxes (and not the derived distances) because that is possible for the FGS parallaxes without any conversions dependent on Bayesian priors, and then the quoted error bars will have a Gaussian distribution.

In a comparison of the {\it HST} and {\it Gaia} parallaxes, we see a relatively good match.  However, the differences are large compared to the claimed error bars.  The worst case is for V603 Aql, where the difference in reported parallaxes is 0.820 mas, while the total 1-sigma uncertainty in the difference is 0.153 mas, so the HST parallax is in error by $\psi$=5.4-sigma.  This is too large to be from random measurement error with the quoted error bars.  And the parallaxes for DQ Her are different at the $\psi$=2.9-sigma level.  This again suggests that the {\it HST} parallaxes have some unrealized and substantial systematic error.  Both of these cases have the {\it HST} parallaxes being {\it larger} than the {\it Gaia} values.  However, GK Per has a $\psi$=-1.3-sigma deviation, while RR Pic has a $\psi$=-0.2-sigma deviation, both in the opposite direction from the first two, with these two cases showing scatter as might be expected from random measurement errors.  So we have a mixed bag for the comparisons, with half the novae having apparent significant systematic errors, while the other half does not.

To be quantitative, the average value of the difference in the distance moduli is $\langle \Delta \mu \rangle=-0.21$, while the RMS scatter of these differences is $\sigma _{\Delta \mu} = 0.37$.  This corresponds to an RMS error in the parallax of 19\%.  The differences in parallax, in units of the total 1-sigma error bar, has an average of $\langle \psi \rangle$=1.7 and an RMS of $\sigma_{\psi}$=3.03.  This goes to show that the FGS parallaxes have systematic errors that average three-times larger than the quoted error bars.  So the real FGS error bars are on average 3$\times$ larger than published.

Unfortunately, the one other {\it HST} FGS parallax to a CV also shows big problems.  In particular, Harrison et al. (1999; 2000; 2004) reported that the prototypical dwarf nova SS Cyg has a parallax of 6.06$\pm$0.44 mas.  But this was found to disagree greatly from the VLBI measured parallax of 8.80$\pm$0.12 mas (Miller-Jones et al. 2013).  Further, the small reported FGS parallax forced SS Cyg to such a high luminosity such that accretion-disk theory strongly states that the dwarf-nova-instability is impossible (Schreiber \& Lasota 2013).  This set up a severe conundrum for the field of CVs (that includes novae).  Most workers in our community thought that the discrepancy was resolved by Nelan \& Bond (2013) when they did a complete reanalysis of the same FGS data and derived a parallax of 8.30$\pm$0.41 mas.  Now, {\it Gaia} gives a parallax of $\varpi_{Gaia}$=8.724$\pm$0.049 mas, and all doubts about the controversy are gone.  So the conclusion is that the early reported FGS parallax must have had some sort of subtle analysis error.  This is disconcerting because the Harrison papers had triply-repeated analysis by the best and most experienced workers in both astrometry and in the {\it HST} FGS.  The lesson from this is that even the {\it HST} FGS parallaxes are sufficiently tricky as to have large systematic errors.

This is all rather discouraging for the FGS parallaxes.  However, to keep the issues in perspective, the FGS nova distances are still the best measures prior to {\it Gaia}.

To further evaluate geometrical parallaxes for CVs, we have a long series of ground-based measures of very nearby systems by J. Thorstensen and coworkers.  In a {\it tour de force}, Thorstensen (2003) and Thorstensen, L\'{e}pine, \& Shara (2008) present 26 geometrical parallaxes for faint and nearby CVs, with the images taken with the 2.4-m telescope at MDM Observatory.  These can be compared to the {\it Gaia} DR2 parallaxes (see Figure1).  I find that the RMS scatter of the differences in distance moduli is 0.54 mag, while the average difference is -0.37 mag.  The differences in units of the total 1-sigma difference have an RMS scatter of 1.06.  With this, we see that the Thorstensen parallaxes have an accuracy only slightly worse than {\it HST}, a moderate bias towards overestimating the CV distances, and accurately reported error bars.  This reliability and accuracy is remarkably good.

\subsection{Testing Expansion Parallaxes From Nova Shells}

Novae often eject visible shells, seen to expand for years and decades.  If we take the expansion velocity of some part of the shell to be given by some part of the wings of the early nova emission lines, then the distance to the nova is a simple calculation from the angular size of the shell and the time since the eruption.  Such distances are called `expansion parallaxes'.  Such distances are known for only around 30 novae.  In the absence of real parallaxes, these distances were perceived as being the best around, and thus the expansion parallaxes became the primary way to calibrate and test other distance methods.

Unfortunately, this method has a greatly-larger real uncertainty than is usually recognized.  (1) The relevant velocity might be given by the Half-Width-Zero-Intensity or the Half-Width-Half-Maximum of the emission line profile, with a factor of two difference in distance.  And many novae have weird `castellated' profiles or P-Cygni profiles, and it is totally unclear as to what to use then.  Indeed, the literature has little discussion and no understanding as to where to pick the velocities from the profiles.  Further, the profiles vary substantially from line-to-line, and the line widths decrease by over factors of two from early-to-late in the eruption.  All of these lead to roughly factor-of-two errors in the distances.  (2) The relevant angular radius can be taken from a wide range of isophotal values, from the peak of some supposed ring, to the edge as defined by the steepest drop in the profile, to the outermost position with shell light.  Many shells are poorly resolved, with some correction required for the point-spread-function of the imaging, with few workers making any such corrections.  Most shells have large-scale out-of-round shapes (with observed axial ratios up to 1.42, see Downes \& Duerbeck 2000), and have knobby features that extend out farther, so which radius should be used?  For the `simple' question of using just one (out of a continuum) of isophotal levels, Wade et al. (2000) have six different ways of defining the effective radius.  The literature has little discussion and no theoretical guidance as to how to choose a radius that corresponds to the velocity somehow selected.  Rather, multiple workers just plea for observers to simply report what they did, a plea that is usually dodged.  Again, these reasonable choices lead to a factor of two uncertainty in the distances.  (3) The radial velocities along the line of sight are usually substantially different from the expansion velocity in the transverse directions.  This is demonstrated by the fact that most novae have substantially out-of-round shells (Downes \& Duerbeck 2000).  This is further shown for many novae with massive high-velocity jets (e.g., GK Per, see Shara et al. 2012b) and for shells with bipolar shapes.  For the third time, we recognize ubiquitous errors at the factor-of-two level, with none of these being discussed much or included in published error bars.

To test the expansion-parallax distances, I have collected published values for 22 novae that are in my Gold and Silver samples.  Most of them have many published values, with these being given in Table 3.  These values have been compiled from Downes \& Duerbeck (2000), \"{O}zd\"{o}nmez et al. (2016), Slavin (1997), and Shafter (1997), which are themselves compilations of values calculated from many sources.  The distances for each nova are not all independent, with many of the results sharing input.

It is disconcerting to see the huge range of values for many of the novae.  This is a strong measure of the large size of the real error bars for the expansion-parallax method.  Further, a third of the published distances have larger than 50\% true errors.  On the plus side, only 12\% of the published distances have errors by greater than a factor of two.

The first set I will evaluate is the distances collected in Slavin (1997), because this includes quoted error bars for all.  For this set of 9 novae, I calculate that $\sigma _{\Delta \mu}$ equals 1.04 mag, corresponding to a 1-sigma uncertainty of a factor of 2.6$\times$ in luminosity.  Further, $\sigma _{\chi}$ equals 3.6, which shows that the reported error bars are on average a factor of 3.6$\times$ too small.  With $\langle \Delta \mu \rangle$= -0.10 mag, there is no bias.  

The next data set is all 75 measures reported in Table 3 (see Figure 2).  Most of these have no quoted error bars.  A measure of the real error in the reported distances is that $\sigma _{\Delta \mu}$= 0.95 mag, which points to an average 1-sigma error of a factor of 2.4$\times$ in luminosity.  The expansion-parallax distances appear unbiased, as $\langle \Delta \mu \rangle$ equals -0.06 mag.  For use in calculating $\sigma _{\chi}$, I have adopted $\sigma_{D_{prior}}=C\times D_{prior}$ as some sort of an average error.  I must adjust $C$ to equal 0.60 so as to get $\sigma _{\chi}$ equal to unity, which is to say that the real error bars are 60\% on average.

What we see is that the expansion-parallax method is far worse than is the popular idea that this is the canonical method.  (But I expect that the researchers who specialize in the method are well aware that the real uncertainties are frustratingly large, c.f. Downes \& Duerbeck 2000; Wade, Harlow, \& Ciardullo 2000.)  The real error is something like 2.6$\times$, and this makes this method largely useless for many applications.  Still, for some applications, the expansion-parallax is the best that we can get, and a factor-of-two is better than no idea at all.

\subsection{Testing Blackbody Distances For The Secondary Stars}

A relatively new method to get a distance to a nova is to derive the blackbody distance to the secondary star.  To do this, we need to isolate the flux from the secondary star and quantify it by its surface temperature ($T$) and a measure of its flux at some wavelength ($f_{\lambda}$).  The radius of the secondary star ($R_*$) is tightly constrained by the orbital period in a Roche lobe filling situation, with little dependence on the star masses.  The luminosity is given by $L=\sigma T^4 \times 4\pi R_*^2$.  For a blackbody spectrum of the secondary, this luminosity can be used to calculate the luminosity at the wavelength of observation, $L_{\lambda}$.  The distance to the nova ($D$) then comes from solving the equation $f_{\lambda}=L_{\lambda}/(4 \pi D^2)$.  The only tricky part is isolating the light from the secondary alone.  Given the hot disk in the system, the secondary can be isolated when it is large and cool, so we can see its near-infrared peak in the system's spectral energy distribution.  Further, to minimize the effects of irradiation on the companion from the hot white dwarf, we need to look at the orbital phase with the unilluminated hemisphere pointing at Earth (hopefully during mid-eclipse).

Few novae have the large cool companion stars required for this method.  Only the RN with red giant companions (T CrB, RS Oph, V3489 Sgr, V745 Sco) and a few novae with orbital periods longer than a day (U Sco and V1017 Sgr) are possible for this method.  Blackbody distances are reported for these stars in Schaefer (2008; 2010), Schaefer et al. (2013), and Salazar et al. (2017).  In practice, this physics-based method promises to be fairly accurate.

The blackbody-distances have proven particularly useful for recognizing an old error that has since become canonized in the literature.  Specifically, one of the first measures of the distance to RS Oph was where Hjellming et al. (1986) claimed a distance of 1600 pc based on a measure of the intervening ISM extinction, with the blunder being that they assumed that the extinction along the entire line-of-sight was that appropriate for the mid-plane of our Milky Way, whereas RS Oph has a galactic latitude of +10.37$\degr$, so the line-of-sight quickly passes outside most of the galaxy's dust and the reported distance is greatly too small (Schaefer 2009; 2010).  This estimate was for a long time the primary published distance, and the later researchers merely cited the 1600 pc distance repeatedly, until the value became canonical and unquestioned (Barry et al. 2006; Schaefer 2009; 2010).  With this canonized error, the system is so close that the calculated blackbody radius of the  secondary star must greatly underfill its Roche lobe (Schaefer 2009).  This then forced unsuspecting theorists to rig models where the accretion onto the white dwarf was entirely by the secondary's stellar wind.  Still, none noticed that it was impossible for such models to get an adequate mass accretion rate onto the white dwarf so as to sustain the frequent recurrent nova events.  It was only when a blackbody distance to RS Oph was calculated that the whole series of blunders came unravelled (Schaefer 2009).  Still, the inertia of all the older astronomers having grown up with the `traditional' distance is a potent bandwagon effect that is hard to overcome.  Into this setting, a {\it Gaia} parallax would settle the distance, even for the old-timers.

Unfortunately, {\it Gaia} DR2 does {\it not} have reliable measures of the parallaxes for the four RN with red giant companions.  The big problem is that the system's binary orbit will make the center of light wobble back and forth with a greater amplitude than the parallax itself.  Further, the orbital periods are comparable to a year, so the {\it Gaia} sampling will inevitably mix and confuse the orbital wobble with the parallax wobble.  Let us take an example of RS Oph, where its orbital period of 453 days (along with stellar masses of 1.3 M$_{\odot}$ for the white dwarf and 1.0 M$_{\odot}$ for the companion), the semi-major axis of the orbit is 1.54 AU, so the orbital wobble has a radius of 0.67 mas at a distance of 2300 pc (Schaefer 2009), which is larger than the parallax of 0.43 mas at that distance.  (Fortunately, the novae with orbital periods of a few days or shorter will have only very small effects on the {\it Gaia} parallaxes.  For an extreme example, GK Per, with its nearby location and a two-day period, should have an orbital wobble of 0.088 mas, with this being greatly smaller than the quoted parallax of 2.263 mas.  However, the orbital wobble is larger than the quoted parallax error of 0.043 mas, so there must be some systematic error introduced by the wobble even if only at the 3\% level.  The only systems where this small effect can be noticed are GK Per and V1017 Sgr.)  So {\it Gaia} DR2 might have reported a formal parallax for the four RN with red giant companions, but these values are certainly wrong due to huge systematic errors arising from their orbital wobbles.

So {\it Gaia} DR2 does not have useable parallaxes for T CrB, RS Oph, V3489 Sgr, or V745 Sco due to their orbital wobble.  And U Sco is too far to produce a useable parallax, and indeed the DR2 parallax is negative.  So the blackbody-distance method can {\it now} only be tested for one nova, V1017 Sgr.  Salazar et al. (2017) give two calculations for the distance by this one method, and concludes that V1017 Sgr is at a distance of 1240$\pm$200 pc.  {\it Gaia} DR2 gives a parallax that translates into a distance of 1269$_{-60}^{+84}$ pc.  The two distances agree closely.  From this one nova as a test for the whole method, I get $\chi = -0.14$ and $\Delta \mu = -0.05$ mag.

For the future, analysis of {\it Gaia} positions with a full model for the orbit will readily produce a very accurate distance and orbit for RS Oph and the other three recurrent novae with red giant companions.  Perhaps {\it Gaia} can also pull out an orbit for AR Cir, which might have a red giant companion, and then a blackbody-distance can be derived and compared.  Another test can be made by using extant observations of GK Per to get its blackbody-distance.

\subsection{Testing Distances From Measures of Interstellar Extinction}

Another ubiquitous method to estimate nova distances is to measure some property of the nova that depends on the column density through the interstellar medium (ISM), somehow know that property as a function of distance, and then spot the distance to the nova.  The measured property is often the color excess, $E(B-V)$, (or one of its variants), or the equivalent width of the ISM sodium absorption lines (or some other ISM line).  The primary problem is always that the function for extinction versus distance has a tremendous amount of scatter.  This problem is notorious, with such derived distances only qualifying as order-of-magnitude estimates.  But many nova researchers (including myself) have often used this method, simply because there is nothing better and we desperately need some sort of a distance.  Now, with the {\it Gaia} DR2 nova distances, we can see exactly how good or bad are the extinction-distances.

Recently, A. \"{O}zd\"{o}nmez and coworkers have pioneered a new method for getting $E(B-V)$ as a function of distance, as based on measures of red clump stars with several infrared sky surveys (\"{O}zd\"{o}nmez et al. 2016; 2018).  Red clump stars have a known absolute magnitude, their reddenings can be measured, and their distances deduced.  This `reddening-distance relation' (RDR) method has a strong advantage that many calibration stars can be measured over a wide range of distances, so the reddening-distance function will have good resolution in distance.  Further, there are many calibration stars near the line of sight to the novae, so the calibration for each novae is for the relevant function with distance.  A disadvantage of the RDR method that some novae are closer or farther than the majority of the extinction so only limits on the distance can be obtained.  Nevertheless, the method has the strong advantage that it provides reasonable distances for most galactic novae.  \"{O}zd\"{o}nmez et al. (2018) has provided an exhaustive list of their derived distances for almost all known galactic novae.  I judge that the RDR method is substantially better than the prior haphazard work.

With the {\it Gaia} distances, we can determine the real accuracy of the RDR method.  \"{O}zd\"{o}nmez et al. (2016) give novae distances for 15 novae that appear in my Gold and Silver samples.  I calculate that $\sigma _{\Delta \mu}$ equals 1.32 mag, $\sigma _{\chi}$ equals 4.25, and $\langle \Delta \mu \rangle$ equals -0.05 mag.  Further, \"{O}zd\"{o}nmez et al. (2018) reports on 5 novae from the Gold and Silver samples, some with distances updated from their earlier work, with these allowing a mostly-independent measure of the accuracy of the RDR method.  For these five novae, $\sigma _{\Delta \mu}$ equals 1.01 mag, $\sigma _{\chi}$ equals 1.31, and $\langle \Delta \mu \rangle$ equals +0.12 mag.  From these numbers, we see an accuracy comparable to that of expansion parallax.  This is a testament to both the nice improvements of the RDR over the earlier haphazard work on the ISM methods, as well as the greatly-poorer-than-advertised accuracy of the expansion parallax method.  Further, we see that the reported error bars are a factor of 1.3--4.2 times too small as compared to their real errors.  And, we see that there is no bias in the reported distances.  In all, we see the RDR as a new method that is applicable to most galactic novae, where its real error bars are comparable to the those of the much-vaunted expansion parallax.

\"{O}zd\"{o}nmez et al. (2018) puts together a catalog for essentially all galactic novae, listing distances and properties, all with good selections.  A fraction of the novae are not able to have RDR distances, usually because the nova is past most of the Milky Way's dust so only a distance-limit is possible.  For novae with trigonometric parallaxes, expansion parallaxes, and black-body distances, these are tabulated instead of the RDR distances.  I calculate for 28 novae in the Gold and Silver samples that $\sigma _{\Delta \mu}$ equals 1.15 mag, $\sigma _{\chi}$ equals 2.45, and $\langle \Delta \mu \rangle$ equals -0.05 mag.  This shows a catalog of nova distances comparable to the accuracy of the expansion parallaxes, with reported error bars 2.45$\times$ smaller than the real error bars, and no bias.  In Section 5, I will be concluding that this catalog of \"{O}zd\"{o}nmez is now the best source for getting novae distances when they are not available from {\it Gaia}.

\subsection{Testing The Maximum-Magnitude Rate-of-Decline (MMRD) Relation}

The MMRD is a relation between the speed of decline and the peak luminosity.  The rate of decline is quantified by either $t_2$ or $t_3$, the time from the date of peak until the light curve has faded by 2 or 3 magnitudes below the peak, respectively.  The peak luminosity is taken to be the absolute magnitude (corrected for extinction) at the maximum in the light curve.  The MMRD relation is the equation that represents the situation that fast-fading novae are more luminous than slow-fading novae.  The best prior versions of the MMRD are from Downes \& Duerbeck (2000), as shown in Figure 4.  Their best fit relation for $t_3$ is $M_{max} = (-11.99 \pm 0.56) + (2.54 \pm 0.35)\times \log [t_3]$.  They concluded that the scatter was so large as to make this useful only for statistical studies.  Despite this clear conclusion, with only poor alternatives, many researchers have applied the MMRD to single systems, often presuming a greater accuracy than is warranted.  Now with the {\it Gaia} distances, we can evaluate the real accuracy of the MMRD, and perhaps tighten the relation.

Historically, the MMRD has been relied on too much (e.g., Schaefer 2010), mainly because it can be applied to most novae, and there are only poor alternatives available otherwise.  In the meantime, a variety of theoretical justifications have been proposed to explain the MMRD (e.g., Shara 1981; Livio 1992).  And the novae in the Large Magellanic Cloud follow the same galactic MMRD (Shafter 2013), again with a large scatter.  Into this situation, Kasliwal et al. (2011) dropped a startling result that the novae in M31 (plus some novae in M81, M82, NGC 2403, and NGC 891) do {\it not} follow the MMRD.  Instead, the M31 novae have a scatter in $M_{max}$ from -6 mag to -10 mag, with few falling anywhere near the Downes \& Duerbeck MMRD.  Further, they showed with my data (Schaefer 2010) on recurrent novae that this subset of Milky Way novae also do {\it not} follow the MMRD (see Figure 5).  This is particularly embarrassing for the MMRD because a substantial portion of many of my distance estimates come from the MMRD, and also because roughly a quarter of the so-called classical novae are actually recurrent novae masquerading as classical novae because only one eruption has been discovered from their multiple eruptions within the last century (Pagnotta \& Schaefer 2014).  Further, Kasliwal et al. (2011) pointed out that theoretical models of nova eruptions (those in Yaron et al. 2005) do {\it not} follow the MMRD.  And then Shara et al. (2017) recently demonstrated that the novae in M87 do {\it not} obey the MMRD, nor any other function, with most all of the novae being far below the galactic MMRD.  With the utter lack of any relation for galactic recurrent novae, M31 novae, M87 novae, and theoretical models of novae, Shara et al. declared in their paper's title that they were ``Snuffing out the Maximum Magnitude-Rate of Decline Relation for Novae as a Non-standard Candle". 

The primary uncertainty for the galactic MMRD has been the nova distances, so now with {\it Gaia} we can test the relation.  To create an MMRD plot (like in Figure 4) for the galactic novae, I have used the light curve data compiled into Table 1 for the Gold and Silver samples.  This is shown in Figure 6.  We see an MMRD with a scatter that is much larger than in Figure 4.  So the better set of distances has not tightened up the MMRD.  What we see is that the prior MMRD relation (solid black line in Figure 4) is a bad and very biased representation of the data.  The scatter is so large that the MMRD cannot even be used for statistical purposes.  Indeed, the scatter is getting so large that we can start questioning whether these galactic nova have a relation at all.  So, the MMRD essentially fails the new {\it Gaia} distances.

Let me quantify this.  For each nova, we can use the Downes \& Duerbeck MMRD relation to get a peak absolute magnitude ($M_{max}$) for the observed $t_3$, and then the distance modulus is the usual $\mu=V_{max}-A_V-M_{max}$.  We get the distance with the usual equation; $D_{prior}=10^{(\mu-5)/5}$.   I am using the Downes \& Duerbeck MMRD as the best of the published relations, because we are testing the prior nova distances derived with this method.  The average total 1-sigma uncertainty for $\mu$ is equal to the observed RMS scatter in the left panel of Figure 4, which equals 0.77 mag.  A plot of $D_{prior}$ versus $D_{Gaia}$ is given in Figure 7.  For the 39 novae in the combined Gold and Silver samples, I calculate that $\langle \Delta \mu \rangle$=0.73 mag,$\sigma _{\Delta \mu}$=1.31 mag, and $\sigma _{\chi}$=1.03.  These show a poor MMRD.  From the start of this study, I took the Gold+Silver sample to be the best for evaluating the prior MMRD distances.  So this poor showing is the best evaluation of the MMRD.

The Bronze sample of 23 galactic novae is defined as those with a confident identification in the {\it Gaia} DR2 catalog for which the fractional error in the parallax ($\sigma_{\varpi}/\varpi$) is greater than 30\%.  By the selection criteria, the Bronze sample must have large error bars in both the distances and parallaxes.  In Figure 8, I plot the prior distances and parallaxes versus the {\it Gaia} values.  Like in Figure 7, the scatter about the diagonal line is huge.  The value parameters for this Bronze set of novae has $\sigma_{\Delta \mu}$=2.20, $\sigma_{\chi}$=1.68, and $\langle \Delta \mu \rangle$=0.44, which are horrible.  But for this sample, the measurement errors are so large that any conclusion about the scatter intrinsic to the MMRD must be weak.  Still, there is some information in the Bronze sample.  However, for most purposes of this paper, the Bronze sample does not provide useful constraints.

Surprisingly, we get two greatly different pictures when we break up the sample into the Gold and Silver groups separately.  For the MMRD plot, this is shown in Figure 9.  We see that the Gold sample is shows an apparently significant relation between $M_{max}$ and $t_3$, albeit with huge scatter and significantly different from the MMRD of Downes \& Duerbeck.  But for the Silver sample, all we see is a scatter diagram with a huge range and the MMRD relation does not exist.  This is a striking difference.  So the rest of section 3.5 will be trying to understand the difference between the two panels of Figure 9, and trying to reconcile them together.

\subsubsection{Possible Selection Effects For Novae Showing the MMRD}

After the utter failure of the MMRD for M31, M87, the RNe, and for theory models, our community was faced with reconciling that some galactic novae nevertheless show the MMRD.  My personal suspicion was that the galactic novae and their measured values were somehow selected out to roughly match the MMRD.  It is easy to consciously-or-unconsciously pick out the distance (e.g., from Table 3) or peak magnitude or extinction such that approximate agreement with the expected luminosity is achieved.  Our nova community should not be offended at this possibility, as such `band-wagon' effects occur too-often throughout physics and astrophysics even in modern times, with notorious and long-lasting cases involving astronomical distances, including the distance to the Large Magellanic Cloud (Schaefer 2008) and the debates over the value of the Hubble constant.  So this possible reconciliation would suggest that some consistent set of input for some sort of a complete and unbiased sample of galactic novae and their properties would show no MMRD, just like for M31 and M87.  

A test of this idea for reconciliation concerns the galactic RNe.  The selection of all ten known systems has nothing to do with validity with respect to the MMRD, and my collection of measured values (Schaefer 2010) was systematic, exhaustive, and independent.  These RNe were selected with identical biases as were the CNe, yet no MMRD is seen (see Figure 5), pointing to the selection effects in Figures 4 and 9a as not being relevant.

I do not see any substantial evidence in favor of a band-wagon effect for the MMRD.  To the contrary, the selection of novae for my SSH sample (i.e., those novae from amongst the 93 best observed novae of all time) was made completely independently of all issues that would affect the MMRD plot.  The novae selected into SSH were picked entirely for the availability of excellent light curve data.  From the SSH sample, our Gold sample was the subset for which {\it Gaia} returns a parallax with $\sigma_{\varpi}/\varpi$<0.3.  Similarly, the MMRD displayed by the LMC novae (Shafter 2013) cannot have selection effects because Shafter was simply taking exhaustive and uniform inputs for all known LMC novae, while the discovery surveys go much deeper than needed to catch all the novae.  So I am not seeing how any bandwagon mechanisms can operate on the Gold and LMC samples.

There is some selection for the Gold sample based on distance and $V_{max}$.  Let us try constructing a selection effect that might create an apparent MMRD.  The MMRD might be manufactured by somehow selecting against slow/overluminous novae and fast/subluminous novae.  Well, the overluminous events will not be selected against, so this cannot account for the dearth of systems well above the MMRD in Figures 6 or 7.  Rather, the void in the upper-right corner of the MMRD plots is apparently real and resulting from the physics of the novae, as confirmed by the void in the upper-right of the MMRD plots for all samples.

So what about selecting against the fast/subluminous novae?  Certainly, the Gold sample has selection for the brighter novae, and this forms a bias against the subluminous events.  But I do not see how this can make for a lack of low-luminosity novae for only the fast events.  Here are my various reasons:  (1) Any selection effects on measured nova properties do not depend on the event duration, so the slow/subluminous novae will have the same bias as the fast/subluminous novae, but this is not seen as many slow/subluminous novae appear in Figure 6.  (2) The discovery efficiency for novae is only a relatively weak function of the peak magnitude (see section 7.1 of Schaefer 2010), the $V_{max}$ distribution in SSH is very broad, and {\it Gaia} returns parallaxes with $\sigma_{\varpi}/\varpi$<0.3 for most novae with quiescent counterparts brighter than around 17 mag.  With this, selection effects can only be weak.  (3) The discovery selection  effects for Silver sample are the same as for the novae in the Gold sample, yet only the Gold sample has no fast/subluminous events.  The discovery selection  effects for recurrent novae are the same as for the novae in the Gold sample (Schaefer 2010), yet only the Gold sample has no fast/subluminous events.  

In all, for trying to reconcile the stark differences between the Gold and Silver samples, I can find no useful evidence or effective logic to attribute the difference to selection or bandwagon effects.

\subsubsection{Possible Data Errors For the MMRD Outliers}

Another possible way to reconcile the two parts of Figure 9 is to simply claim that the outliers in the Silver sample are erroneous.  After all, the Gold sample is the best data, so it is not surprising that a sample of inferior data could have far outliers that mask the underlying MMRD.  This is an easy and glib way to defend the MMRD.  And this reconciliation is plausible.  However, it is a dangerous route for scientists to glibly ignore a large fraction of the data so as to defend some traditional and long-used idea.  Rather, we should closely examine all the inputs ($V_{max}$, $A_V$, $t_3$, and $D_{Gaia}$) to see if we can impeach the result so as to allow the nova to fit the MMRD.  In this section, I will consider the eight novae that are outliers from the MMRD in Figure 6.

	{\bf CI Aql:}  CI Aql has well-measured $V_{max}$, $t_3$, and $A_V$ values, while the identity of the quiescent counterpart in the {\it Gaia} DR2 results is of high confidence.  The only way that I can think of that CI Aql can be impeached for placement in the MMRD plot is to declare that recurrent novae are somehow different from classical novae and the MMRD is not expected to apply.  But any such {\it post facto} creation of an exception to get rid of one outlier is bad science.  More specifically, the physics of classical novae and recurrent novae are identical, so the MMRD should apply to the entire range of eruptions.  Further, the RNe form a continuum with the classical novae (where the recurrence time scale spans a broad range), so there is no reason to think that the nova mechanism has a sudden change at any specific recurrence time scale.  And about a quarter of the so-called classical novae are really RNe with multiple eruptions within the last century (Pagnotta \& Schaefer 2014), so many novae in our Gold sample should also be outliers, but such is not seen.  In all, CI Aql is a very confident outlier to the MMRD.

	{\bf BC Cas:}  BC Cas is a sparsely observed nova, so we can glibly think that the true values for the input are sufficiently different from those in Table 1 so that agreement with the MMRD might be reached.  Well, the confident light curve in Duerbeck (1984) might be sparse, but it is adequate to give a more-than-good-enough measure of the maximum magnitude.  That is, there is a pre-peak limit that tightly constrains the peak, and color effects are too small to make a difference.  To get to the MMRD, the peak would have to be 2.5 mag brighter than observed, and this is not plausible.  The extinction value from Harrison, Campbell, \& Lyke (2013) is reliable, and $A_V$ is certainly not 2.5 mag larger.  Further, Liu \& Hu (2000) has confidently identified the quiescent counterpart, and it is certainly the source listed in the {\it Gaia} catalog.  In all, there is no way to impeach the placement of BC Cas on the MMRD plot, so it remains a far outlier.

	{\bf AR Cir:}  AR Cir has a poorly observed light curve, and it is not impossible that the quiescent counterpart is the g=20.3 star a bit further northwest of the `bright star'.  But the best way to impeach its placement on the MMRD plot is to note that AR Cir might well be a symbiotic nova.  (A symbiotic nova eruption is distinct from a classical nova eruption that happens to occur in a CV binary where one component is a red giant.  The real symbiotic nova eruptions have light curves that have durations of a year to a decade or more and low amplitudes from 1 to $\sim$5 mags.  A CV accreting system with a red giant will formally be a symbiotic star, having both a hot and cold component in their spectrum, but a thermonuclear eruption on the white dwarf  can be either a normal nova or a greatly-different symbiotic nova.)  The evidence for this is a low outburst amplitude, a nominal $t_3$ of 330 days or longer, and an unresolved companion of late type (Harrison 1996).  However, the amplitude of 8 mag is too large for a symbiotic nova eruption, so this identification of AR Cir as a symbiotic nova is problematic.  The physics of symbiotic novae is different from that of classical novae, so we have no reason to think that they should follow the MMRD.  (Similarly, we should not be placing Type Ia supernova onto the MMRD plot.)  Thus, by pushing past the evidence (in particular the 8 mag amplitude), we might think that AR Cir is not useful to be an example of an extreme outlier from the MMRD.

	{\bf V1330 Cyg:}  V1330 Cyg only lies 1.5 mag off the Downes \& Duerbeck MMRD.  The nova was discovered on 8 June 1970 (with no useful pre-discovery plates), at which time the light curve was already slowly fading.  Ciatti \& Rosino (19) have spectra to show that the nova was 20--25 days past peak at discovery, putting the extrapolated peak around 15 May 1970 at an estimated 7.5 mag.  This extrapolated peak is 2.4 mag brighter than tabulated in SSH and Table 1, and this would be enough to bring V1330 Cyg into agreement with the MMRD.  But Ciatti \& Rosino also estimate $t_3 \sim$ 20 days, apparently as based on the observed decline rate starting a bit after peak.  (The fast decline is also given by the He/N nature of the spectrum, as well as by the high expansion velocity.)  Further, they measure that the $B-V$ color is always zero or negative, so we must have $A_V \sim$ 0.0 mag.  With these changes to Table 1, we have $M_{max}$= -4.8 mag (for a fast $t_3$) and V1330 Cyg is 3.9 mag below the MMRD.  So while attempts to impeach the inputs reveals large uncertainties, V1330 Cyg does appear to be a far outlier.

	{\bf BT Mon:}  BT Mon is 2.8 mag below the MMRD.  This nova has a well-observed light curve with a flat maximum lasting $>$60 days (Schaefer \& Patterson 1983), with this serving as the prototype of the F-class for nova light curves (SSH).  The spectral evidence places the time of maximum around the time of the start of the flat maximum.  It is possible for a willful researcher to speculate that there was a peak $\sim$2.8 mag brighter and before the observed flat maximum, but such an unprecedented light curve shape can only be adopted by someone desperately trying to save the MMRD from another outlier.  So I conclude that BT Mon is a confident outlier.

	{\bf HZ Pup:}  HZ Pup is 3.0 mag below the MMRD.  The light curve of Hoffmeister (1965) shows a maximum extending $>$58 days, jittering up and down, with a deep limit 21 days before the maximum.  There is no real chance that the the light curve could have a significantly higher maximum crammed into the 21-day interval.  The $t_3$ value is certainly long, while the extinction from Harrison et al. (2013) is good.  The identification of the quiescent counterpart is certain, and this is confidently matched to the {\it Gaia} DR2 catalog source.  There is no plausible way to impeach the data, so HZ Pup is a confident outlier to the MMRD.

	{\bf V1016 Sgr:}  V1016 Sgr lies 1.8 mag below the MMRD.  The light curve is sketchy, but just enough information is available to be confident that the basic parameters are reasonably measured (Pickering 1910).  The first positive detection was on 10 August 1899 at 8.5 mag, whereas on the previous night it was fainter than 11.5 mag, and the nova faded from 8.6 mag on 25 August to 10.5 mag on 13 October.  This is enough to give a maximum of close to 8.5 mag, and $t_2$=64 days.  The subsequent slow fading can be interpolated to give $t_3$=140 days.  \"{O}zd\"{o}nmez et al. (2018) gives $E(B-V)$=0.35$\pm$0.04, while Shafter (1997) closely agrees.  The {\it Gaia} parallax refers to a fourteenth magnitude star at the correct position (Duerbeck 1987).  The only weak link that I can see is that I know of no spectroscopic or photometric proof that the fourteenth magnitude star is the real quiescent counterpart, rather than some fainter star that was not recognized by {\it Gaia}.  While this scenario is possible, there is no positive evidence against the common identification, so any such attempt to impeach the {\it Gaia} distance can only be wishful-thinking speculation, at least for now.  So V1016 Sgr appears to be a good outlier of the MMRD, but the final proof of the identification is not known.

	{\bf V721 Sco:}  V721 Sco is the farthest outlier of the MMRD, being 6 mag below the fit from Downes \& Durebeck.  The nova was discovered first by G. Haro on 3 September 1950 at 9.5 mag, faded fast to 11.7 mag on 8 September (when F. Zwicky independently discovered the nova), and continued fading to 13.0 mag on 12 September (Herzog \& Zwicky 1951).  For pre-discovery images, the Palomar 48-inch telescope showed no star to 18.0 mag on 16 August (18 days before the first discovery).  Table 1 and Figure 6 have adopted a peak of 8.0 mag, as evaluated by Shafter (1997).  The peak magnitude cannot be greatly brighter than the discovery magnitude of 9.5, or else the eruption could not fit into the 18 day interval.  Further, after 12 September, the rate of fading slowed substantially, so the transition is apparently around that date, with the peak-to-transition amplitude being $\sim$4 mags, for a peak of near 9.0 mag.  With this new evaluation, the discrepancy with respect to the MMRD only becomes worse, at 7 mag.  There is no chance that the peak was greatly brighter than Shafter's 8.0 mag.  And the $t_3$ value is definitely short, much under 10 days.  Harrison et al. (2013) gives $A_V$=1.1 mag, while Shafter (1997) gives $A_V$=2.4 mag, with it being impossible for the extinction to be so large as to make any difference.  Further, the {\it Gaia} catalog entry is for the star in the corner of an `inverted L' at exactly the coordinates given by Duerbeck (1987) as based on a Harvard A plate showing the star in eruption.  So the only way to try to impeach the input for V721 Sco is to assume that the real quiescent counterpart is much fainter than this star, and so close to its position as to be unrecognized by {\it Gaia}.  But making such an evidenceless speculation is just circular (assuming that which is being sought).  So I conclude that V721 Sco is indeed a strong example of a very-far outlier for the MMRD.
	
	In all, with eight far outliers to the MMRD from our Gold and Silver samples, none have been impeached with enough confidence to change Figure 6.  Indeed, six of these eight novae are highly confident as being outliers.  For the other two novae (V1330 Cyg and AR Cir), the best evidence is that they are outliers.  In only one case (AR Cir), do we have a possible reason to remove the outlier, and that is to push past the amplitude limit for symbiotic nova eruptions, declare that the eruption must have been a symbiotic nova (as opposed to the system being a symbiotic star because it has a red giant companion), and then presume that the MMRD does not apply to symbiotic novae.  What all this is saying is that this section's attempt to reconcile the different results from the Gold and Silver samples (as simply being measurement errors) has failed completely.

\subsubsection{Possible Differences In Populations}

	Perhaps Figures 9a and 9b are different because the Gold and Silver samples largely consist of novae from two separate populations, with the MMRD applying to one of those populations but not the other.  Similarly, we can speculate that the MMRD applies to the LMC novae of one population, but that the MMRD does not apply to the M31 and M87 novae of some different population.
	
	My first idea was that the Gold and Silver samples might be dominated by either populations in the bulge or the disk, and perhaps the MMRD is applicable to only one of these populations for some unknown reason.  But this reconciliation does not work because the Gold sample, the Silver sample, and the outliers are nearly all from a disk population.  This is inevitable, as {\it Gaia} produces useable parallaxes for the quiescent novae out to $\sim$3000 pc, and so they must all be $\gtrsim$5000 pc from the galactic center, so the disk population must dominate.  Further evidence for the dominance of the disk population is that the mean value of $\cos(\Theta)$ (with $\Theta$ being the angle between the nova and the galactic center) is 0.13, 0.17, and 0.26 for the Gold, Silver, and outlier samples, respectively.  And the three samples have appropriate concentrations towards the galactic plane, as expected for disk populations.  The LMC novae of Shafter (2013) are all of a young population, while the novae of Kasliwal et al. (2011) are mostly far from the bulge in spiral galaxies.  The M87 novae of Shara et al. (2016; 2018) must all be of something like an older bulge population, as M87 is an elliptical galaxy.  The galactic RNe are evenly divided between thick disk and bulge populations (Schaefer 2010).  So we see that the Gold and Silver samples are both of the same population, and there is no consistent story as to whether a young disk population will have the MMRD applicable.
	
	Another set of divisions amongst the novae relates to the light curve classes (SSH).  These are `S' for smooth light curves, `J' for light curves with large flares around the time of peak, `D' for novae showing significant dust dips soon after the maximum, `P' for smooth light curves that have a plateau around the transition phase, `O' for novae showing periodic oscillations around the transition phase, `C' for novae with a prominent rebrightening with a slow rise and fast fall in a `cusp' shape, and `F' for smooth light curves that have the maximum being a long-lasting flat top.  We could imagine that the MMRD might apply to only some of these classes, and not apply to others.  However, the eight outliers include classes P, S, F, and J; spanning from the more-energetic to the less-energetic events.  Further, the Gold, Silver, and Bronze samples have consistent distributions across the classes, spanning many classes.  So the difference between the samples does not appear to be related to the light curve classes.
	
	From Figure 5, we see that the recurrent novae do not follow the MMRD, while Pagnotta \& Schaefer (2013) show that a quarter of the `classical novae' are really recurrent.  So maybe the real classical novae (i.e., with recurrence time scale greater than a century) are the ones for which the MMRD is applicable, while the true RNe (with the recurrence time scale faster than one century, whether observed or unobserved) are the ones for which the MMRD is not applicable.  But this does not work, as the Gold sample has 2 RNe, 2 candidates, and 22 CNe, while the Silver sample has no RNe, 4 candidates, and 6 novae that are certainly CNe (Pagnotta \& Schaefer 2014).  And of the outliers, there is only one RN, no candidates, and 5 certain CNe (Pagnotta \& Schaefer 2014).  That is, the novae that fall along the MMRD relation are both RNe and CNe, while the novae far from the MMRD relation are a similar combination of RNe and CNe.  So the RN/CN division cannot reconcile the Gold and Silver samples.
	
	So this entire attempt to reconcile the Gold and Silver samples has failed, as there is no apparent population difference between the samples, nor are the outliers significantly of a separate population.  This bodes bad, because we have no means to distinguish a nova as to whether the MMRD is applicable.  For any nova without a {\it Gaia} parallax, we cannot know whether the MMRD applies to give a bad distance estimate, or whether it is inapplicable and giving a random distance estimate.

\subsubsection{Conclusions Concerning The MMRD}

From the preferred Gold+Silver sample, we see a poor MMRD with a large scatter, substantially offset from the prior MMRD, and with 20\% of the novae as far outliers.  The errors in the MMRD are up to 6 mags, while 7\% of the novae have >2.5 mag errors in the distance modulus.  This is unacceptable for all applications.  The real scatter is so large that the MMRD (if it exists at all) is useless for studies of individual novae and even useless for statistical purposes.  From this, I conclude that the MMRD has completely failed as a prior method to get nova distances.

I am purposely not giving any `best fit' MMRD based on the new {\it Gaia} distances.  This is because then we might have incautious workers being tempted to use any such new relation. 

We still have the question as to the very existence of a relation between the rate of decline and the $M_{max}$.  The problem is that two data sets (the Gold sample and the LMC novae) show a significant yet poor relation, whilst five data sets (the Silver sample, the galactic RNe, M31 novae, M87 novae, and grids of theoretical light curves) show that there is no such relation as an MMRD.  After considering selection effects, bandwagon effects, data errors, and various population differences, I found that none could reconcile the differences in the applicability of the MMRD.  So this question remains unresolved.

The MMRD has a huge scatter for novae for which it apparently applies, and for most novae the MMRD does not apply at all.  We cannot distinguish which novae the relation applies, so it is a bad bet to apply any MMRD to any individual nova or statistically to any sample.  So my advice is that our community must relinquish prior distance estimates based on the MMRD, and all the resulting conclusions.  Further, I advise that our community no longer use or publish any distance estimates from the MMRD, even if it is the only distance estimate available. 

\subsection{Testing  the Constant-$M_{15d}$}

Buscombe \& de Vaucouleurs (1955a; 1955b) pointed out that novae have a nearly constant absolute magnitude at a time of fifteen days after the peak of $M_{15d}=-5.2\pm0.1$, with this applying to both Milky Way and LMC novae.  van den Bergh \& Younger (1987) used six galactic novae (with distances from expansion parallaxes) to find $M_{15d}=-5.23\pm0.39$.  Schematically, this can be see by over-plotting many V-band light curves shifted right-left so all the peaks line up and shifted up-down such that the magnitudes are all corrected to be absolute magnitudes.  Ideally, all the light curves then roughly cross each other at a time of 15 days after the peak.  The cross-over is at $M_{15d}=-5.23\pm0.39$, and can serve as a standard candle.  This is essentially an alternative formulation of the MMRD, as a fast nova with a small $t_3$ will have a high-luminosity $M_{max}$, while a slow nova with a large $t_3$ will have a low-luminosity $M_{max}$.

Shara et al. (2018) used ten weeks of daily HST imaging of the giant elliptical galaxy M87 to get great light curves of 41 novae so as to test the constancy of $M_{15d}$.  (This is the same data set used to refute the MMRD in M87 by Shara et al. 2017.)  They find a weak convergence of light curves towards latter times.  With the F606W filter (a bit redder than the V band), they find an average $M_{15d}$ of -6.37 mag, with an RMS scatter of 0.46 mag, after they select out 16 novae with decline times $t_2>$10 days.  With the F814W filter (between the R and I bands), they find an average $M_{15d}$ of -6.11 mag, with an RMS scatter of 0.43 mag, from 17 not-fast novae.

I have tested the distances derived by the constant-$M_{15d}$ method.  For 34 novae in the Gold+Silver sample, I calculate that $\sigma_{\Delta \mu}$=1.53 mag (showing poor accuracy), $\sigma_{\chi}$=2.26 (showing that the prior method underestimates the real error bars by 2.26$\times$), and $\langle \Delta \mu \rangle$=0.04 mag (showing no bias).  As a variation on the MMRD, this method returns just as poor a distance, with huge scatter.  In particular, for the Gold and Silver samples, the total range of the absolute magnitude at a time fifteen days after peak is from +1.6 mag (for V721 Sco) to -7.4 mag (for V732 Sco).

I have further tested the selection for not-fast novae with $t_2>$10 days.  For 23 novae in the Silver and Gold samples, $\sigma_{\Delta \mu}$=1.08 mag, $\sigma_{\chi}$=2.21, and $\langle \Delta \mu \rangle$=-0.28.  This selection of non-fast novae is a bit of an improvement.  But the accuracy is still poor, and the error bars are a factor of 2.21$\times$ too small.  The total range of $M_{15d}$ is from -2.8 mag (for HZ Pup) to -7.4 mag (for V732 Sco), with this huge range meaning that the method is useless.

\subsection{Testing  the Constant-$M_{max}$}

So far, other than the parallaxes, all the prior nova-distance methods have been poor.  And the previously-canonical methods ({\it HST} parallax and the expansion parallax) could be applied to few novae.  For the MMRD methods (including the constant-$M_{15d}$ method), detailed light curve information is required, and such is not available for the majority of the novae.  Often enough, the only useful information is the peak magnitude, so it would be nice if only that were required to get a distance.  So we could consider a nova-distance method where we only assume that the peak absolute magnitude is some constant, then calculate the distance modulus and the distance.  Crudely, novae come to the near the same luminosity, so this `method' should not be horrible.  Still, I have never seen this method published anywhere, likely because the existence of the claimed MMRD relation makes this method seem naive.  But now, we see that the MMRD often apparently does not exist, so this constant-$M_{max}$ method is the only estimate that can be used for all the faint poorly-observed events.  This method does have poor utility, provided that appropriately large error bars are acknowledged.

Another use of this constant-$M_{max}$ method for this paper is that it serves as some sort of a null hypothesis to compare the other nova-distance methods.  That is, we can see that any methods that approach the accuracy of this naive simplistic method are not worthy of use.  A better statement of this is that the use of extra fit parameters (like $t_3$) are not justified by any significant improvement in the fit.

With this, I will take a round number value of $M_{max}$=-7 mag as the basis for this naive method.  For this, I will take the 1-sigma uncertainty in $M_{max}$ to be 1.4 mag, selected so that the error bars reflect reality of the {\it Gaia} distances.  I can then calculate the various quality measures for 37 novae in the Gold+Silver sample.  I calculate that $\sigma_{\Delta \mu}$=1.61 mag (showing poor accuracy), $\sigma_{\chi}$=1.00 (the value is unity by construction), and $\langle \Delta \mu \rangle$=0.00 mag (so my round number shows no bias).  For the Gold and Silver samples, the total range (see Figure 6 and 9) is from -3.5 mag (for AR Cir) to -9.5 mag (for CP Pup).  This total range and RMS scatter is poor, so the $M_{max}$=-7 mag method will produce errors too large to be useable for most purposes.  

A comparison of the $\sigma_{\Delta \mu}$ values for the best MMRD and the constant-$M_{max}$ methods is 1.31 mag versus 1.61 mag, with the difference being comparable to the uncertainties.  Still, the prior MMRD distances are greatly worse than the naive method, because the $\sigma_{\Delta \mu}$ value only measures the scatter around some best fit line, whereas the MMRD has a systematic offset of 0.73 mag, so the actual errors in the MMRD distances from the prior best MMRD relation will be greatly larger.  So the best MMRD is worse than the naive empirical description that the peak luminosity is a round-number constant.  This is saying that the peak absolute magnitude has no significant variation with $t_3$, which is to say that the MMRD is false.

In the final section, I will be concluding that the $M_{max}$=-7 mag method is the last resort for those faint novae with no {\it Gaia} parallax and no distance from interstellar extinction.  That is, nova-distances are sometimes needed for the faint novae, and no better information may be available than by somehow using the observed $V_{max}$.  The constant-$M_{max}$ method is comparable to the MMRD, yet the constant-$M_{max}$ method does not have the aura of sophistication, physics, or accuracy that is wrongly carried along with the MMRD.  That is, when some researcher uses the MMRD, readers might expect that they get something better in accuracy and reliability than they actually get.  Whereas, a simple use of the constant-$M_{max}$ method will not fool anyone into thinking that we have anything better than an empirical description of a range, with no physics and only poor accuracy.  Critical to the use of the $M_{max}$=-7 mag method is that a realistic error bar be attached.  The 1-sigma on $M_{max}$ is 1.4 mag and the total range is from -3.5 to -9.5 mag.  The error bar must go along with any such derived distance, as without an explicit statement, a reader might go away thinking that the derived distance is more accurate than it really is.  So the $M_{max}$=-7.0$\pm$1.4 mag method is poor, but for many faint novae, it is the only useful means to get a reliable distance.

\subsection{Testing Combined Distances}

Many papers discussing individual novae consider and combine distances from a variety of sources and methods, coming to some sort of a middle value.  This has the advantage that multiple measures should improve the accuracy, and that poor results get moderated.  

To illustrate and test these combined measures, I will here report on my published combined distances for novae that {\it Gaia} DR2 provides useful distances.  (1) For V1017 Sgr, Salazar et al. (2017) reported six estimates, for which only the blackbody-distance to the secondary star measure had any reliability, to conclude that the distance is 1240$\pm$200 pc.  This is to be compared favorably with the {\it Gaia} distance of 1269$_{-60}^{+84}$ pc.  (2) For BT Mon, a combination of the MMRD and extinction measures led to a final adopted distance of 1500 pc (Schaefer \& Patterson 1983).  Now, the {\it Gaia} distance is very close, at 1477$_{-84}^{+128}$ pc.  (3) For T Pyx, I reviewed (Schaefer 2010) prior results from many methods and agreed with Patterson that the distance is 3500$\pm$1000 pc.  Three years later, Schaefer et al. (2013) noted that the literature reports claimed distances ranging uniformly from 1000 pc to $\geq$4500 pc, while a critical reanalysis concluded that the distance was anywhere from 1000 pc to 10,000 pc.  As a note added after the original submission, we agreed that the Sokoloski et al. (2013) measure of the light sweeping through the light echo gave a reliable distance of 4800$\pm$500 pc.  Into this rich history of many workers with many measures, {\it Gaia} now gives the real distance to T Pyx to be 3185$_{-283}^{+607}$ pc.  So the middle ground proves to be correct, while the unique light-echo-distance is just over 2-sigma off.  (4) For IM Nor, Schaefer (2010) gave 3400$_{-1700}^{+3400}$ pc, as based on extinction and the MMRD, with the quoted error bars being huge.  For {\it Gaia}, IM Nor is in the Bronze sample, for a distance of 1205$_{-119}^{+2116}$ pc.  With both measures having large error bars, the good overlap is not meaningful.  For calculating $\sigma{\Delta \mu}$, this one distance measure dominates because my quoted value is nearly a factor of three off, despite being well within the quoted error bars.  (5) For CI Aql, Schaefer (2010) decided to not use model-dependent theory measures, which gave distances around 1500 pc, but instead chose to give the average from three versions of the MMRD, 5000$_{-2500}^{+5000}$ pc.  {\it Gaia} gives 3189$_{-315}^{+949}$ pc.  We see consistency, although the error bars are large, while the model-based distances are poor.  For these five novae, I get $\sigma _{\Delta \mu} = 0.93$ mag, $\sigma _{\chi} = 0.86$, and $\langle \Delta \mu \rangle = 0.82$.  The majority of the bias and scatter in $\Delta\mu$ just come from IM Nor, for which I quoted a huge error bar.  These value measures show that my combined estimates still have substantial error (comparable to that for the expansion parallaxes), a substantial bias (mostly from the one IM Nor distance estimate that had an admitted large uncertainty), and reasonable error bars.

H. Duerbeck provided distances for 35 novae as ``based on nebular expansion parallaxes, interstellar line strengths, differential galactic rotation, and several other methods"  (Duerbeck 1981).  Duerbeck was one of the most experienced and broad workers on novae, and I have long trusted his judgments.  His set of combined distances includes 17 novae in the Gold and Silver samples, all of which are in the Gold sample.  He quotes error bars for only 8 of these novae.  I calculate that $\sigma _{\Delta \mu} = 1.05$ mag, $\sigma _{\chi} = 4.38$, and $\langle \Delta \mu \rangle = -0.43$.  This shows substantial real errors (comparable to that for the expansion parallaxes), a moderate bias (towards underestimating the distances), and error bars that are greatly too small (by a factor of 4.38$\times$).

J. Patterson put together a very influential and comprehensive paper that collected and evaluated many nova distances from many methods (Patterson 1984).  Patterson had the good judgement and the very broad knowledge to put together consensus distances for many cataclysmic variables with up to nine distance clues.  He gave distances for nine classical novae that are represented in the {\it Gaia} Gold+Silver sample.  Just looking at the ratios of Patterson's distances to the {\it Gaia} distances, I find the RMS scatter of the differences in distance moduli ($\Delta \mu$) to be 0.42 mag, and a near-zero average.  Patterson does not (and could not) quote formal error bars.  For calculating the $\chi$ values, I have made the assumption that Patterson's error bars are all a constant fraction of his quoted distance.  If Patterson is assumed to have a 16\% error on his quoted distances, then $\chi$ has an RMS of unity, while the average is 0.02.  So Patterson has a 1-sigma error of 16\% (0.42 mag in the distance modulus) and has no bias long-or-short.  So with $\sigma _{\Delta \mu} = 0.42$ mag and no bias, Patterson has managed to do roughly as good as {\it HST}, long before its launch.

Patterson et al. (2013) provides a list of CV distances as based on multiple methods.  This includes 11 dwarf novae (of the ER UMa subclass) and four classical novae, of which 13 are found with reliable parallax measures in the {\it Gaia} DR2 catalog.  I calculate that $\sigma _{\Delta \mu} = 0.44$ mag and $\langle \Delta \mu \rangle = +0.17$.  No individual error bars are quoted on Patterson's distances, although he suggests that ``errors are probably 25-35\%".  With my formulation that all the errors are a constant fraction of the {\it Gaia} distances, I find that $\sigma _{\chi} = 1$ for $C=17$\%.  This is the only case where the real error bars are substantially {\it better} than quoted in print.  Patterson again has the accuracy of the {\it HST} parallaxes, all with no significant bias.

\subsection{Testing Distances Estimated With Theoretical Models}

It is possible to develop a theoretical model specific for an individual nova system so as to calculate some sort of a luminosity, which when compared to the observed brightness will give a distance modulus.  Such model distances are occasionally given, and often further quantities are calculated from the distance, sometimes as a test of the model.  Such distances really should not be used for most purposes, because we do not want applications built on ever-shifting theory.

Theory-model-distances have a poor history.  Here, I will detail the cases for novae that I happen to have worked on, mostly reported in Schaefer (2010):  (1) Already in Section 3.8, I have told about the theoretical models for CI Aql that give distances around 1500 pc, while the {\it Gaia} distance is 3189$^{+949}_{-315}$ pc, for over a factor of two error.  (2) Occasionally, workers have claimed that novae cannot exceed the Eddington limit, place the peak luminosity near the Eddington limit, and get a distance from that.  The Eddington limit corresponds to an absolute magnitude of roughly -7.0 (Selvelli et al. 2008) or -6.75 (Duerbeck 1981).  But Kasliwal et al. (2011) found that 80\% of the novae in M31 violate the Eddington limit by 2$\times$ to 16$\times$.  Further, from the Gold+Silver novae with well-observed peaks, 9 systems peak more luminous than -8.0 mag, so super-Eddington peak flux is common.  So we now know that a substantial fraction of ordinary novae do exceed the limit (after all, they are explosively expanding), so these old theory-distances are wrong by a large factor.  (3) For U Sco, the model-based distances have a wide and time-variable range, even for one group of theorists.  We read distances of 3300--8600 pc (Kato 1990), 4100--6100 pc (Hachisu et al. 2000a), 6000--8000 pc (Hachisu et al. 2000b), and 6700$\pm$670 pc (Hachisu \& Kato 2017).  This situation is made worse because {\it Gaia} DR2 has a negative parallax for U Sco and a 1-sigma lower limit on its distance of 14,300 pc.  So this set of model-distances from one group covers a wide range, variable over time, and more than a factor of two in error.  (4) For RS Oph, the original Hjellming mistake (claiming 1600 pc from their data) has been endlessly repeated, forcing modelers to have the mass transfer be by a stellar wind, and then these models were used to confirm the Hjellming mistake (Barry et al. 2006).  (5) Theory has provided models to explain the MMRD, and hence supports these distances.  But the MMRD has completely failed, demonstrating that the theoretical support was dubious.  (6) For V394 CrA, Hachisu \& Kato (2000) get a model distance, but they used $E(B-V)$$\cong$0.0 during the eruption and at the same time used $E(B-V)$=1.10 during quiescence, with no reasonable physical mechanism to explain such a huge difference.  Such large and critical internal inconsistencies leave a poor impression for model-distances.

As a representative test of model-distances, I can take a series of papers by I. Hachisu and M. Kato (Hachisu \& Kato 2016a, 2016b, 2018).  They have created a `generalized color-magnitude diagram for nova outbursts' where the light curves follow a `universal decline law' with free-free emission dominating the spectrum.  They calculate luminosities from their model and then derive the distances.  They report such distances for 26 novae from Table 1 (16 in the Gold sample and 10 in the Bronze sample).  The 1-sigma error bars variously are 10\% or 20\%, depending on the light curve.  I calculate that $\sigma_{\Delta \mu}$ is 0.76 mag, although the inclusion of the Bronze sample returns a value of 1.66 (with several far outliers).  The $\sigma_{\chi}$ value is 1.66, pointing out that their claimed error bars should be 1.66$\times$ larger on average.  The $\langle \Delta \mu \rangle$ value is 0.16 mag, pointing to a nearly unbiased case.  Given the substantial variations in the light curves past what their model can address, I judge their moderate $\sigma_{\Delta \mu}$ value and their unbiased $\langle \Delta \mu \rangle$ to represent a success for their model as making largely-reasonable predictions for the {\it Gaia} distances.

\subsection{Collecting  Quality Measures for Prior Methods}

My calculated values for the three quality parameters for various sets of novae and for all the prior nova-distance methods are collected into Table 4.  For most of the lines, the quality parameters are $\sigma_{\Delta \mu}$, $\sigma_{\chi}$, and, $\langle \Delta \mu \rangle$.  But for the first two lines involving parallaxes, the quality value is $\sigma_{\psi}$ instead of the closely similar $\sigma_{\chi}$.

The number of nova in each set is $N_{nova}$.  With only one nova, $\sigma_{\Delta \mu}$ and $\sigma_{\chi}$ are undefined, with the single $\Delta \mu$ recorded in the $\langle \Delta \mu \rangle$ column and the single $\chi$ value reported in the $\sigma_{\chi}$ column.  When $N_{nova} \lesssim$ 5, the statistics are usually dominated by the largest outlier.  

The $\sigma_{\Delta \mu}$ value is telling us the RMS scatter of the errors of the prior method's derived distance modulus, and this is the primary measure of the quality of the prior method.  Ideally, we want this value to be small.  The range for $\sigma_{\Delta \mu}$ is set by the {\it Gaia} parallaxes at 0.24, all the way up to the value for the naive null-hypothesis method ($M_{max}$=-7) with 1.61.

The $\sigma_{\chi}$ value is really telling us about the reliability of the size of the quoted error bars for the prior method.  If the value is around unity, then the reported error bars are reliable in their size.  The $\sigma_{\chi}$ value directly tells us the factor by which the published error bars need to be increased to be reliable.  So for example, on the first line of Table 4, the {\it HST} parallaxes have $\sigma_{\chi}$=3.03, which implies that the average published error bars are nearly 3$\times$ too small.

The $\langle \Delta \mu \rangle$ value is a measure of the bias (high-or-low) in the distance modulus from the prior published values.  Ideally, the value should be near zero, but any value within something like a quarter of a magnitude of zero implies no substantial bias to within measurement accuracy.  For cases with low $N_{nova}$, for example for the nova light echo method, the large values are just random jitter and likely do not point to a real overall bias in the method.

\section{Conclusions On Prior Nova Distances and Methods}

Now, we can compare and contrast all the nova-distance methods:

Pre-{\it Gaia} geometric parallax remains the best of all the prior methods.  That is, the prior parallaxes have small $\sigma_{\Delta \mu}$ and no bias.  The ground-based based parallaxes of Thorstensen have accurate error bars, but the {\it HST} parallaxes have reported error bars that are on average 3$\times$ too small.  Unfortunately, only four novae have any prior parallaxes, and these are now superseded by the new {\it Gaia} parallaxes.  Further, there will be no more nova parallaxes from the ground or from {\it HST} because all the nearby novae are done, and there is no way to compete with {\it Gaia}.

The expansion parallax method has long been the canonical nova-distance method, but this reputation is not deserved.  The prior distances are consistently reporting error bars that are a factor of 3.6$\times$ too small on average.  The better workers in the field know of these problems, and they point to an handful of sources of errors, each of which is roughly 2$\times$ uncertainty.  But the end users of the expansion parallaxes are taking the published error bars at face value, ignoring the reality of the large systematic errors.  Nevertheless, the real indictment of the expansion parallax method is that $\sigma_{\Delta \mu}$ is close to 1.0 mag.  The fractional error in the distance will be $\sigma_D/D=10^{\sigma_{\Delta \mu}/5}$, or the average 1-sigma error in distance will be from a factor of 1.6$\times$ too small to a factor of 1.6$\times$ too large.  This means that the luminosity and energetics will have a 1-sigma range from 2.5$\times$ too small to 2.5$\times$ too large.  For many purposes, such a large uncertainty is not useful.  Further, expansion parallax distances are only known for moderately near novae, and the distances for these are now better known from {\it Gaia}, so these are superseded by the new results.  Historically, another primary purpose of the expansion parallax has been statistical in calibrating other nova-distance methods, but this application is now past, as {\it Gaia} distances have supplanted it for calibration.  For the future, the expansion parallax method will only have utility (with an acknowledged realistic error bar) for recent novae for which the quiescent counterpart is too faint for {\it Gaia}.  This is a rather small area of poor utility for only a few individual novae.  So I am seeing the whole expansion parallax method as now only one of historical interest, for which the real accuracy is greatly worse (by 3.6$\times$) than advertised in the old literature.

The blackbody-distance-to-the-secondary-star method proved quite accurate in the one case that can be tested with DR2.  This is promising for this physics-based method.  This method is also promising because the published error bars for six novae have $\sigma_D/D$ between 16\% and 23\%.  But the real test of this prior method can only come in the near future when {\it Gaia} produces positional fits that includes the orbital motions.  The primary limitation of this method is that it can only be applied to novae with orbital period of longer than a day or so, and this is just a half-dozen systems, mostly recurrent novae.  In the future, there can be few additional applications, so this method has little further utility.

The interstellar-extinction distances were widely known to be horrible due to the very large scatter of extinction measures as a function of distance over nearby lines of sight.  Nevertheless, the literature is full of such estimates, largely because researchers were desperate for even an order-of-magnitude distance.  Into this situation, \"{O}zd\"{o}nmez et al. (2016; 2018) have found a way to improve the measure of extinction as a function of distance along the nova's sightline, and they have produced a systematic catalog of distances to most novae.  Their distances are unbiased, but their error bars are on average too small by a factor of 2.45$\times$.  Critically, their novae distance have $\sigma_{\Delta \mu}$=1.14 mag.  This is poor, yet it is nearly the same accuracy as the much vaunted expansion parallax.  Still, this method has nice utility for the future, because it it the best method for all the many novae too far away for {\it Gaia} to get better than a limit on the distance.  Indeed, this seems to be the only reliable method for the many novae too faint or too far for {\it Gaia}.

The MMRD is the worst of all the prior methods.  The {\it Gaia} distances prove that the prior MMRD distances are greatly biased ($\langle \Delta \mu \rangle$=0.73) and hugely scattered ($\sigma_{\Delta \mu}$=1.31 mag).  This real one-sigma error bar for the prior published distances, with their large systematic offset from the prior MMRD actually makes the prior published errors bigger than what we have for the naive $M_{max}$=-7 method.  And the $M_{15d}=-5.23$ version of the MMRD is only worse.  And the MMRD does not apply at all to most novae samples.  So our community should forego any use of the MMRD.

The naive $M_{max}$=-7.0$\pm$1.4 method has the striking advantage that it can be applied to most of the faint and distant novae, where better methods cannot be applied.  Another striking advantage is that this method is easy, and the input ($V_{max}$ and $E(B-V)$ to moderate accuracy) is readily known for many faint novae.  Another striking advantage is that this method does not carry the aura of false-accuracy, physics-input, or wrong-sophistication, such as might be attached to the MMRD relations, so no one is fooled.  The striking disadvantage is that the error bars are large, so large that the resultant distances may be too poor in accuracy for many applications.

Many papers on individual systems try to combine information from many methods, with this presumably being an improvement on any one method.  It is like crowd sourcing of many poor values making for a better result.  The results from papers by H. Duerbeck and myself still have substantial real errors, comparable to that of the expansion parallaxes.  However, distances of Patterson (1984; 2013) are unbiased, have only 16\% real error bars, and $\sigma_{\Delta \mu}$=0.42 mag.  So Patterson was able to use his good judgement plus many methods to get distances as good as the {\it HST} parallaxes.  Unfortunately, it will be difficult to transfer Patterson's knowledge and skill to other workers.

Theoretical modeling of measured system properties to get a distance is bad for many reasons.  First, for many applications, the use of theory will just be circular reasoning.  Second, we do not want observational conclusions based on any theory, as such is too `flexible', and later building would then be on an ever-changing basis, with the origin soon lost.  Third, the method has a long history of making large errors.  Estimating distances from theoretical models is a method that should not be used.  

\section{Recommendations for the Future}

Nova distances are needed for a variety of applications, so what have we learned from {\it Gaia} as a guide to future practices?  Well, the easy and obvious recommendation is to simply use the latest {\it Gaia} distances.  (This is only for systems with confident quiescent counterparts that have certain {it Gaia} detections, and only when $\sigma_{\varpi}/\varpi <30\%$, all using the EDSD Bayesian prior.)  But this works for only perhaps the nearest 20\% of systems.  So I should make recommendations on how to proceed for the many nova with no useable {\it Gaia} distance.

The recommendations and predictions for the future depend on the needed task.  For calibrating various relations, there are enough good-quality {\it Gaia} distances and luminosities (i.e., the Gold+Silver sample) so as to supersede all other sources of distances.  For nova statistics defining the astrophysics of the systems, the obvious path is to start with the Gold+Silver sample, and if more novae are needed, then to use the catalog of \"{O}zd\"{o}nmez et al. (2018).  Nova have in the past been of high importance for getting the galactic distance scales and the Hubble constant, but now, even with the {\it Gaia} results, novae are no longer competitive in this era of `precision cosmology'.  Nova distances are still vital for modelers of individual nova systems, with the best path depending on the details.

For individual novae, the most tempting path for secondary researchers is to simply troll through the literature and reach some conclusion for follow through.  For the fainter and more-distant novae with no {\it Gaia} distances, the only available method with any merit is the interstellar-extinction of \"{O}zd\"{o}nmez et al. (2018).  When applying these catalog results, I would recommend increasing the quoted error bar by a factor of 2.45$\times$.  For further advances on this catalog, values of $E(B-V)$ must be collected and decided upon, with novae usually have various measures of wide dispersion.  Presumably, the researcher will avoid the trap of somehow selecting out one preferred value (for example, their own measured value), but will instead combine all measures in some appropriate manner, hoping to emulate Patterson.  I would advise and insist that realistic error bars, including statistical errors, be used in the weighted averages.  Now we know from {\it Gaia} that most prior published error bars are greatly too small, and the naive use of incorrect error bars can only lead to poor science.

A trap for researchers is to include MMRD distances in their evaluations for individual novae.  This bad temptation might be based on the bad-science that they should use the prior MMRD distances because the values are published and hence are sacrosanct.  The MMRD-temptation also can arise if there is no other useful distance information.  Now, we know that the MMRD has uselessly-huge real error bar.  Now we know that the MMRD applies only to some small fraction of all novae, and we have no way of knowing if any individual nova is in that small fraction, so it would be past reckless to use this method.  So I strongly recommend that MMRD distance from the prior literature (e.g., Table 27 of Schaefer 2010) be excised and corrected, and never used again.  Further, I strongly recommend that future workers should never put forth any MMRD distance.

This still leaves unanswered what a researcher should do if the nova does not have a distance in the catalogs of {\it Gaia} or \"{O}zd\"{o}nmez et al. (2018).  A glib answer is that such novae should not be used for any purpose needing a distance.  Nevertheless, many researchers will be tempted to somehow use some additional information and get some crude distance.  The galactic coordinates can put some likely very-loose constraints, while assuming $M_{max}$=-7 mag is reasonable provided the large uncertainty is explicitly included.  But there is no other method of any useable confidence that can be applied to the faint galactic novae.

So let me give a cookbook for recommended procedures to get the best distances for galactic novae:  (1) If a confident quiescent counterpart can be certainly identified in the {\it Gaia} database, and there is a quoted parallax with $\sigma_{\varpi}/\varpi <$30\%, then use the distance calculated from the parallax with the EDSD Bayesian prior.  (2) If no reliable {\it Gaia} parallax is available, then look in the catalog of \"{O}zd\"{o}nmez et al. (2018).  Do not forget to multiply the quoted error bars by 2.45$\times$.  For new novae or for trying to improve cataloged distances, it is fair to measure the $E(B-V)$, hopefully with multiple methods, and then construct a judicious weighted average, with realistic error bars, for insertion into the red-clump-stars method for getting extinction as a function of distance.  (3) If no useable distance is found, then take the easy and empirical method that $M_{max}=-7.0\pm1.4$ mag.  For this path, you must explicitly highlight the real error bars.  That is, the 1-sigma error bar is 1.4 mag, while the total range, from Table 1 for the Gold+Silver sample, is from -3.5 to -9.5 mag.  (4) If there is not enough light curve information to get $V_{max}$, then there is nothing more that can be done of any useable merit, and the nova should not be used for any purpose where a distance is needed.


{}

\begin{figure*}
	\includegraphics[width=\columnwidth]{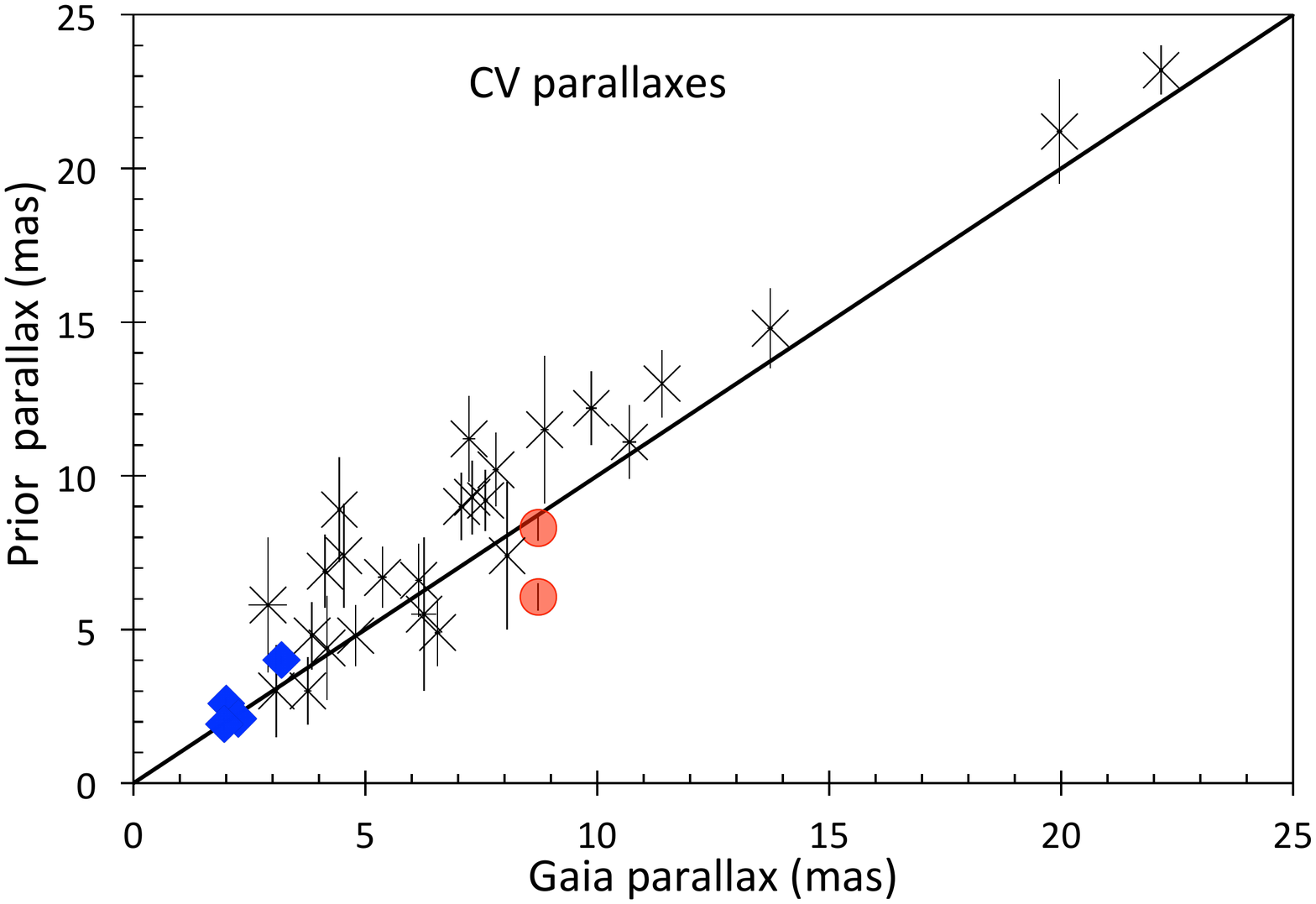}
    \caption{Prior parallaxes for novae and cataclysmic variables (CVs).  This plot shows the prior parallaxes and the {\it Gaia} parallaxes for four novae (blue diamonds), SS Cyg (red circles), and nearby cataclysmic variables ($\times$s).  The diagonal line is to guide the eye for $\varpi_{Gaia}=\varpi_{prior}$.  The nova and SS Cyg measures have the prior parallaxes with the {\it HST} FGS, and these show real errors that are 3$\times$ larger than their quoted error bars.   Despite these problems, the {\it HST} parallaxes for just four novae are by far the best distances prior to {\it Gaia}. The ground-based parallaxes from Thorstensen and coworkers for many CVs were made with a 2.4-m telescope, with their parallaxes having comparable accuracy (i.e., $\sigma_{\Delta \mu}$) with realistic error bars.}
\end{figure*}

\begin{figure*}
	\includegraphics[width=\columnwidth]{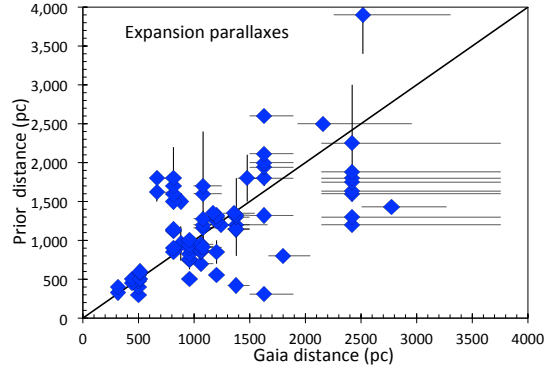}
    \caption{Expansion parallaxes for novae.  Prior expansion parallax distances have been published for 22 novae that appear in my Gold and Silver samples, often with many published widely-varied distances for each nova.  As a test of the prior expansion parallaxes, this plot shows the distances versus the {\it Gaia} ground-truth.  We see a horrifyingly large scatter.  But for the last many decades, expansion parallaxes were considered as the canonical method.}
\end{figure*}

\begin{figure*}
	\includegraphics[width=\columnwidth]{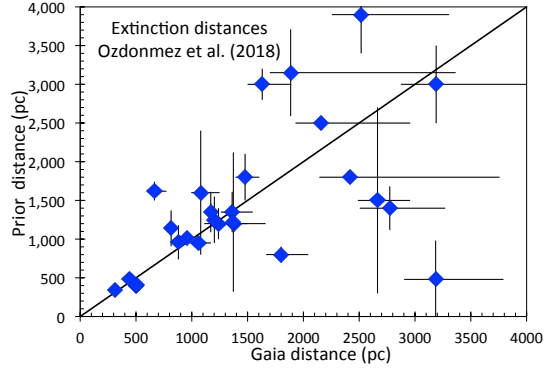}
    \caption{Testing nova distances based on extinction.  The prior nova distances from \"{O}zd\"{o}nmez et al. (2018) are based on many independent measures of $E(B-V)$ for each nova, with the extinction as a function of distance along each sightline calibrated with red clump stars.  This exhaustive and consistent work has improved on the earlier hodge-podge of estimates because they are looking closely at the nova's line of sight, they used consensus extinction values, and their distances are uniformly evaluated for most novae.  This plot shows a large scatter, but that the extinction-distances are not biased.}
\end{figure*}

\begin{figure*}
	\includegraphics[width=\columnwidth]{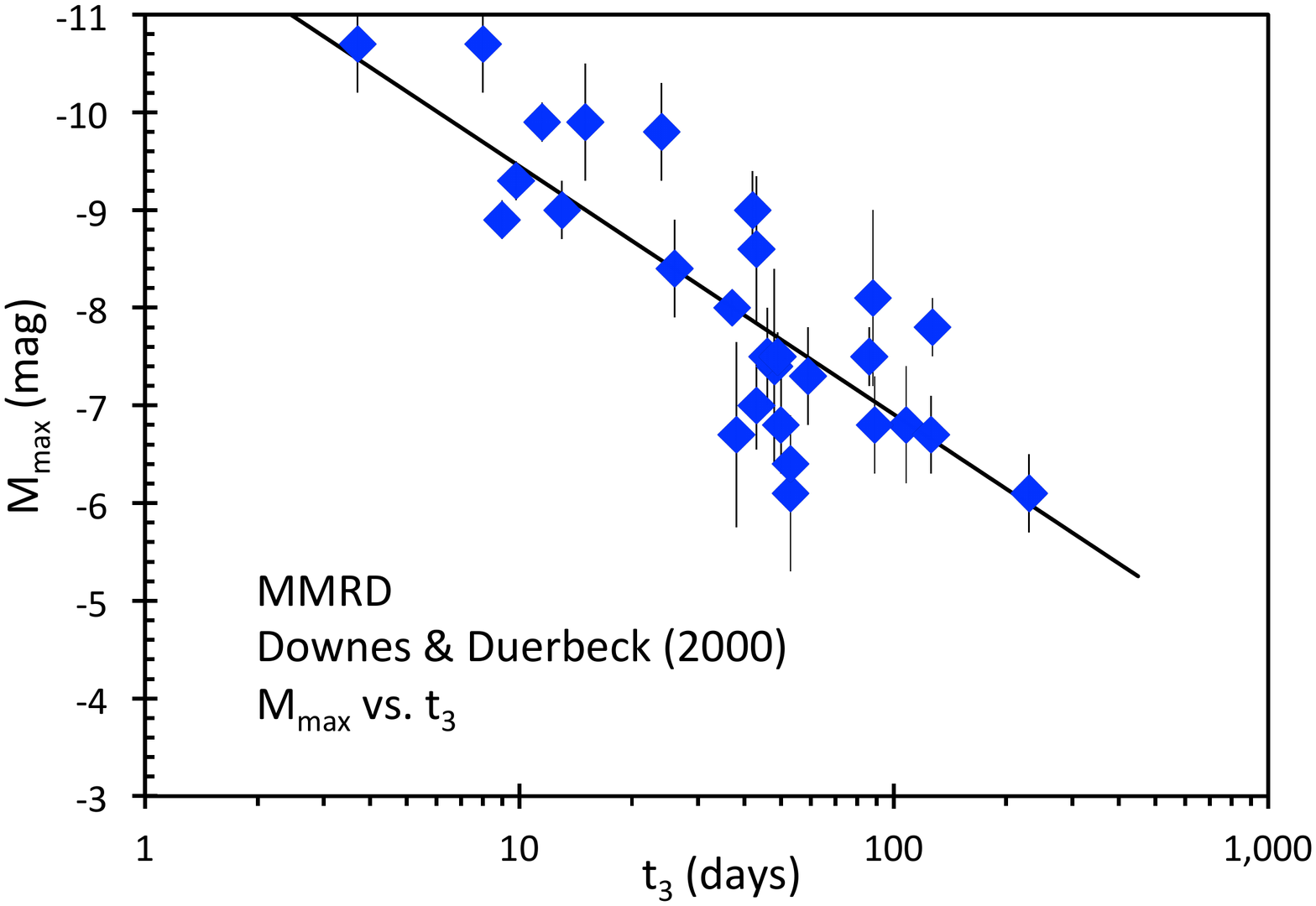}
	\includegraphics[width=\columnwidth]{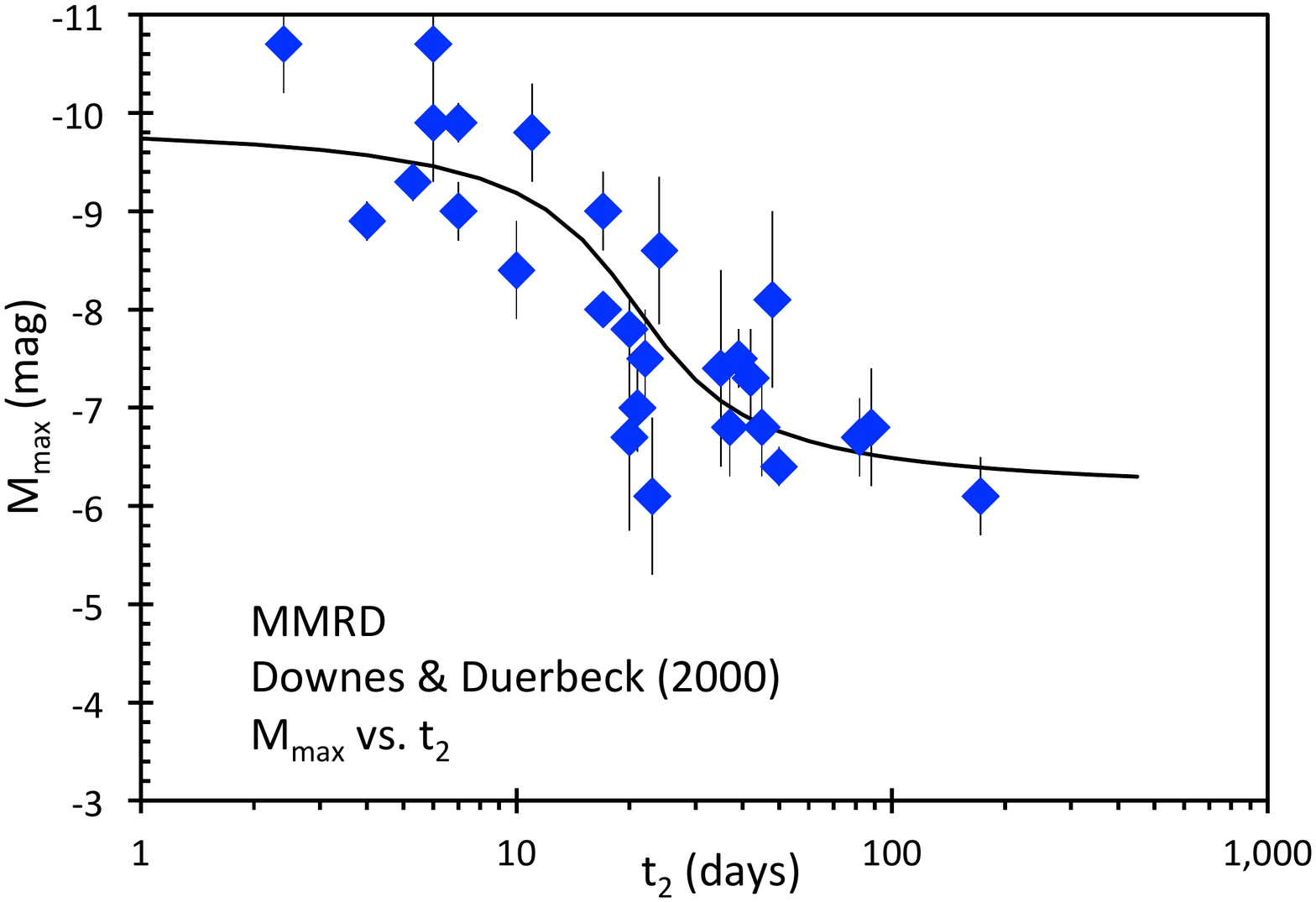}
    \caption{The prior MMRD relation.  Here, I have plotted the maximum magnitude versus rate of decline relation as given in the exhaustive Downes \& Duerbeck (2000).   The maximum magnitude is expressed as the absolute magnitude at peak in the V band, $M_{max}$.  The rate of decline is quantified by the number of days it takes the nova light curve to decline from peak to 2.0 mag below peak ($t_2$) or to 3.0 mag below peak ($t_3$).  The left panel plots the observed values for $t_3$ versus $M_{max}$ as blue diamonds from Table 5 of Downes \& Durebeck, along with their best fit straight line.  The right panel is for the $t_2$ version of the relation, along with the four-parameter $\arctan$ function as the best fit.  It is this diagram that was used to calibrate the MMRD, as applied over the last two decades.  With these plots, we see that this collection of Milky Way novae do appear to definitely show the MMRD relation, even though the scatter is huge.  That is, it appears that the faster-fading novae are more luminous.}
\end{figure*}

\begin{figure*}
	\includegraphics[width=\columnwidth]{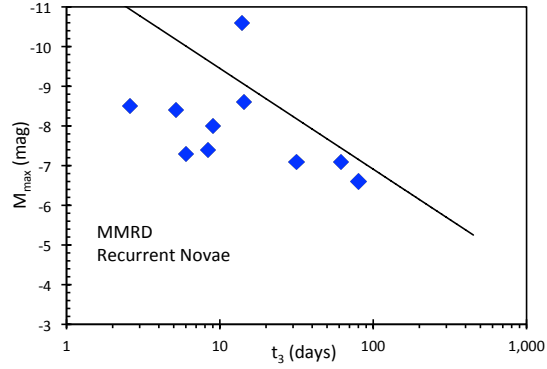}
    \caption{The MMRD for recurrent novae.  If we select out the galactic novae for which more than one eruption has been discovered in the last century or so, then we really expect to get an MMRD similar to the plots in Figure 4.  But we see that (1) the RN do not follow the Downes \& Duerbeck MMRD, (2) the scatter is huge about any relation, and (3) there is no significant correlation between $M_{max}$ and $t_3$.  This says that the MMRD does not exist for this sample of galactic novae.}
\end{figure*}

\begin{figure*}
	\includegraphics[width=\columnwidth]{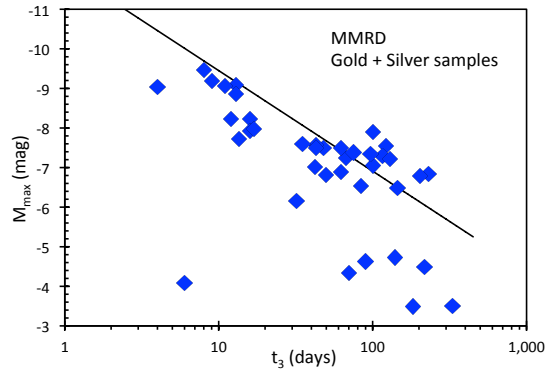}
    \caption{The MMRD for the Gold+Silver samples of novae.  These 39 novae constitute a test of the MMRD from Downes \& Duerbeck (the black line).  The MMRD fails badly.  Most of the novae are below the line, so the MMRD relation is greatly biased.  Critically, the scatter is huge, so that the use of any function will return distances that are greatly in error.  Thus, the use of the MMRD is bad for any purpose (individual or statistical) in past or future papers.  The scatter is so huge that we have to call into question the very existence of any MMRD relation (where fast novae are more luminous than slow novae).  Nevertheless, a desperate defender of the old ideas can willfully make a {\it post facto} tossing out a quarter of the data, and the remaining points look similar to the large scatter in Figure 4, albeit with a greatly different best-fit equation.  So, on the face of it, this plot shows that the MMRD does not apply to galactic novae in any useful way.}
\end{figure*}

\begin{figure*}
	\includegraphics[width=\columnwidth]{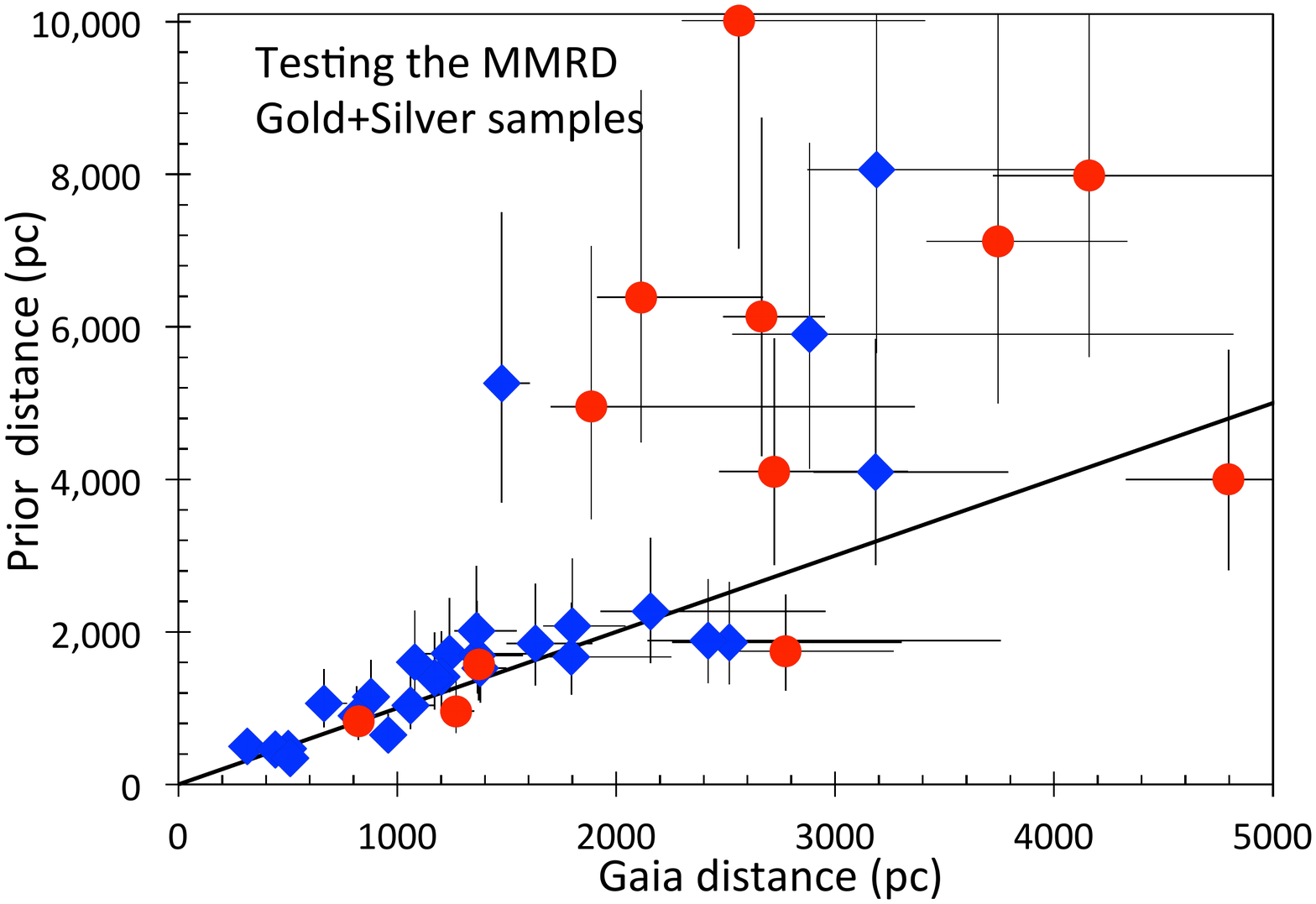}
	\includegraphics[width=\columnwidth]{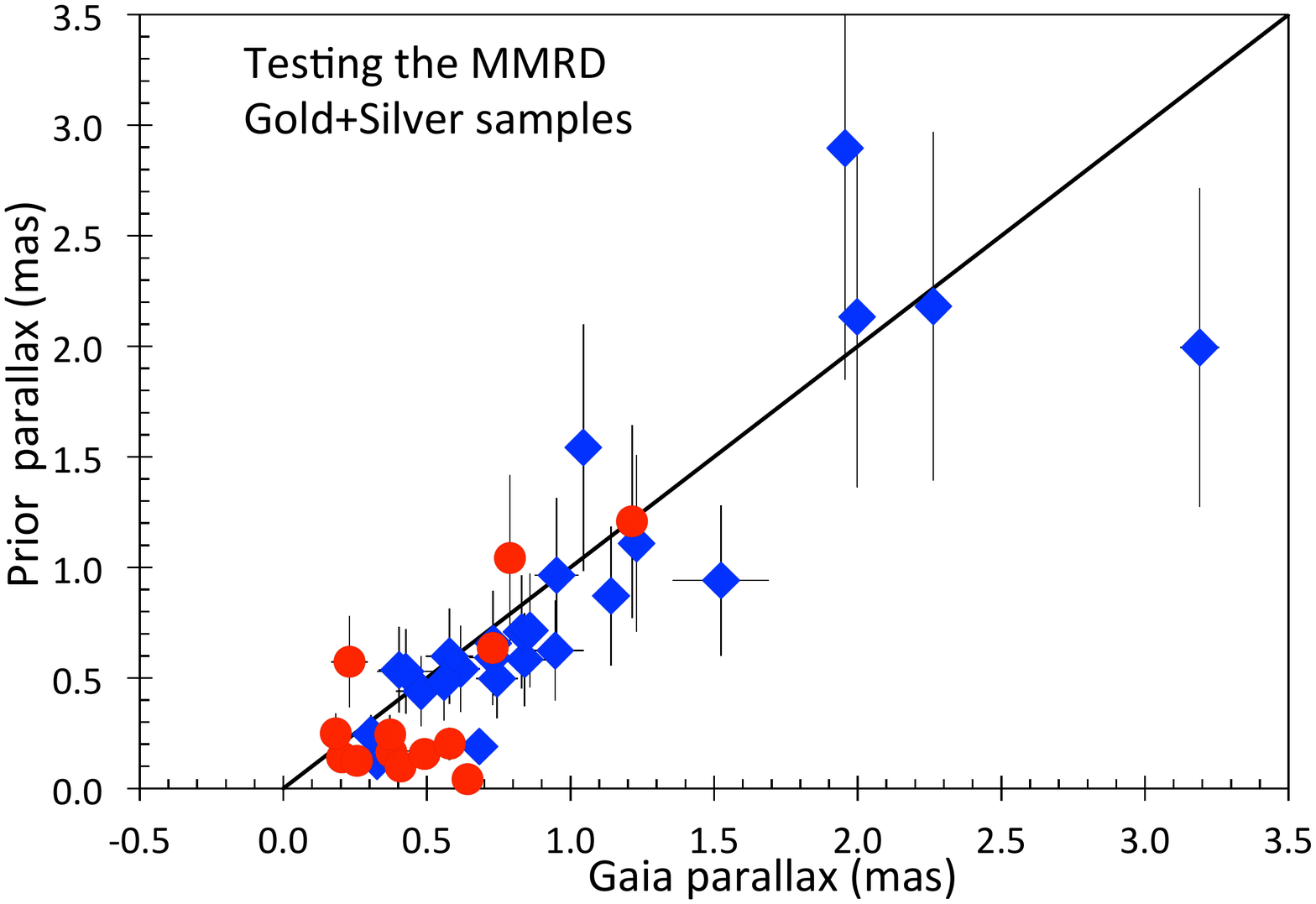}
    \caption{The distances and parallaxes from the MMRD.  These two panels both show the MMRD-derived values for the Gold sample (blue diamonds) and the Silver sample (red circles).  These are essentially the same data as given in Figure 6.  The plot of $D_{prior}$ versus $D_{Gaia}$ on the left side is an easy way to see the accuracy and errors of the MMRD.  If the MMRD is good, then the points should cluster along the diagonal line to within the plotted 1-sigma error bars.  We see immediately that the MMRD has many novae with huge errors.  (Indeed, there is one Silver-sample nova, V721 Sco, that did not fit onto the plot for any useful range, being at $D_{prior}$=$24100^{+10300}_{-7200}$ pc and $D_{Gaia}$=$1570^{+160}_{-100}$ pc.)  We see 8 out of 39 novae have $>$2$\times$ errors in distance (or $>$4$\times$ in luminosity).  Further, with 29 out of 39 novae having $D_{prior}>D_{Gaia}$, we see that the old MMRD has a large systematic bias.  Another way of looking at these same data is to plot $\varpi_{prior}$ versus $\varpi_{Gaia}$, as in the panel on the right.  Again we see that most of the novae are below the $\varpi_{prior}=\varpi_{Gaia}$ line (shown as the black diagonal line), many of the points deviate by more than a factor of two from the diagonal line, and 46\% of the novae are $>$1-sigma from the line.  All this goes to show that the MMRD is too poor in accuracy to use for individual novae or to use for statistical purposes.}
\end{figure*}

\begin{figure*}
	\includegraphics[width=\columnwidth]{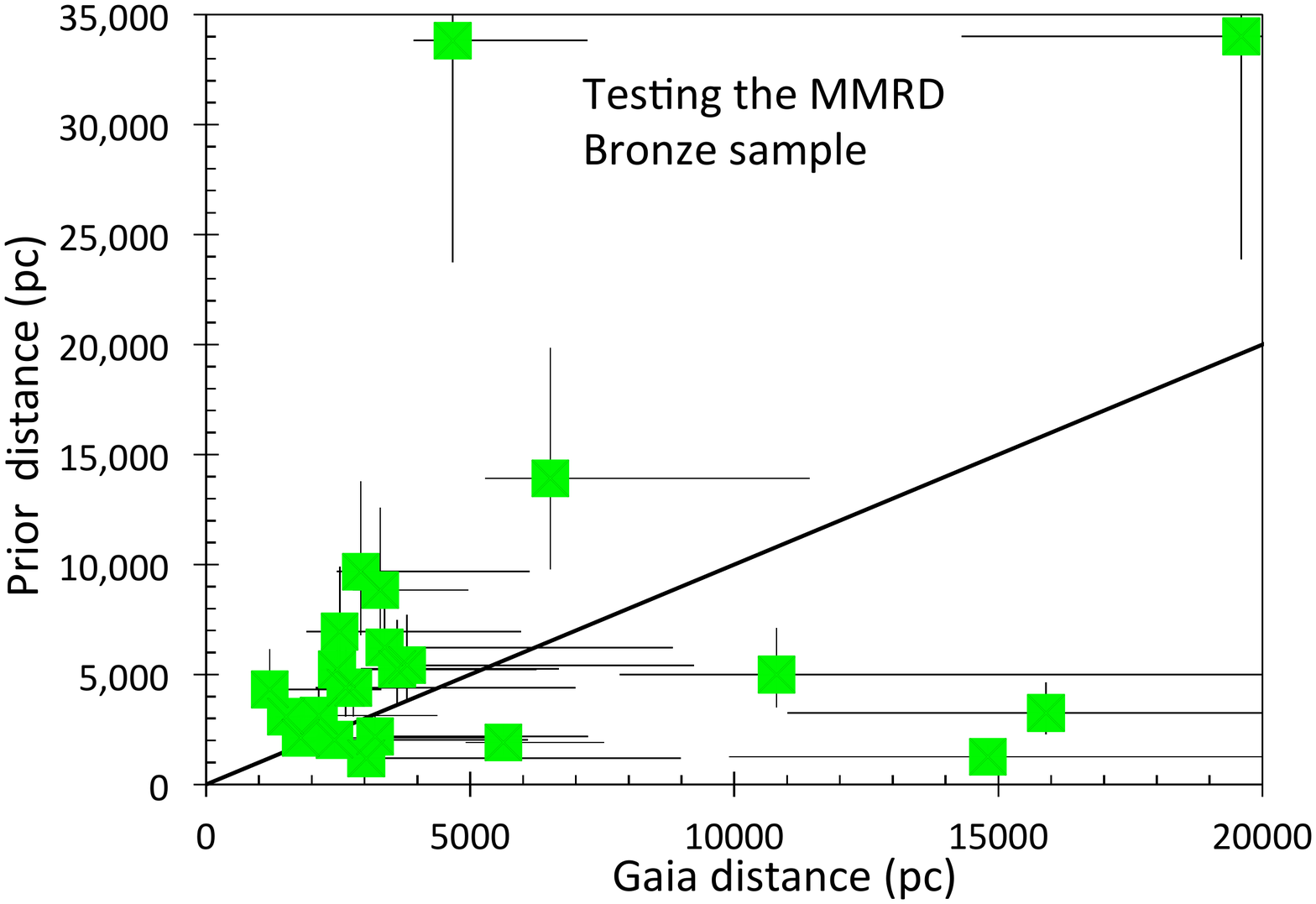}
	\includegraphics[width=\columnwidth]{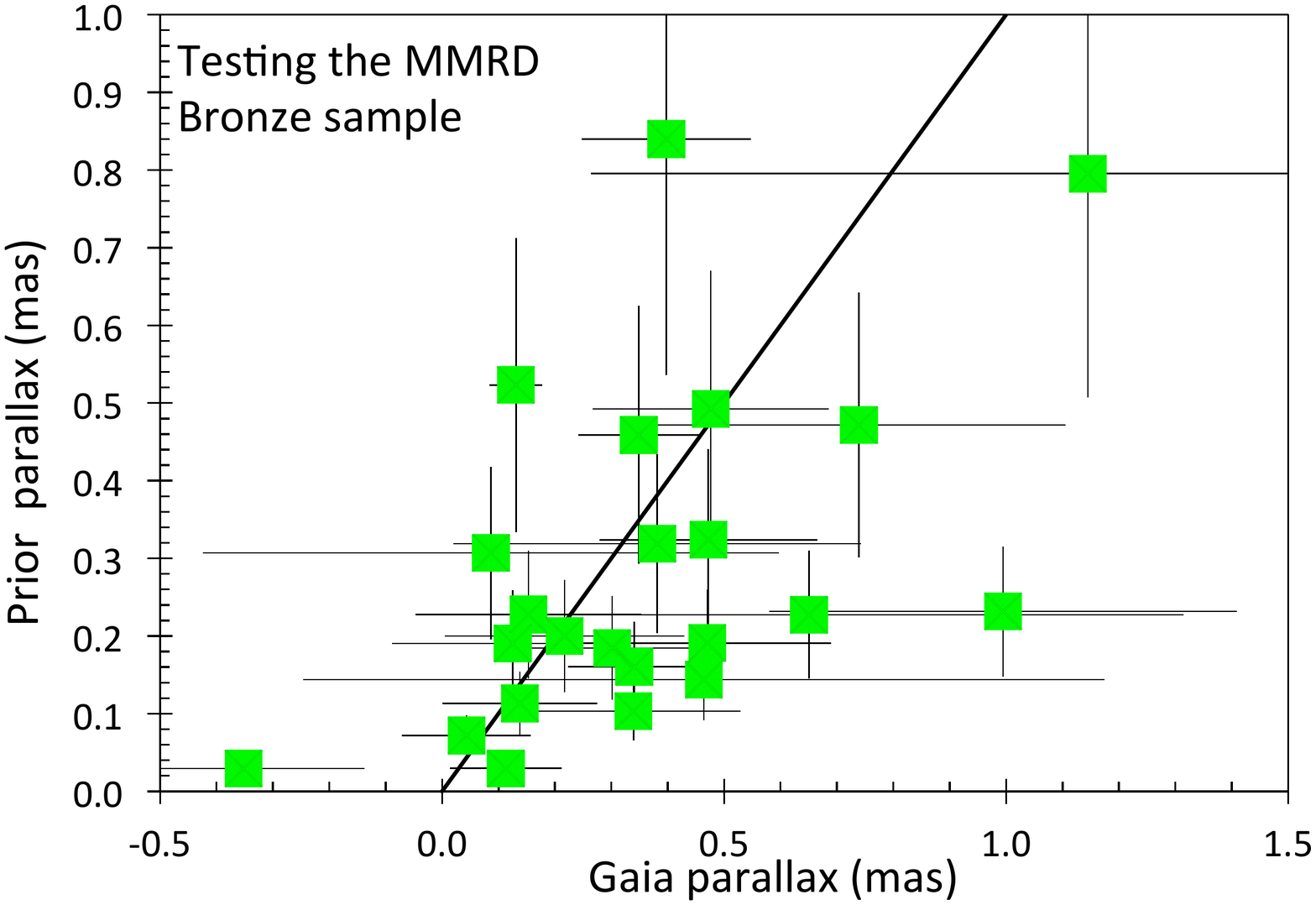}
    \caption{Testing the MMRD with the Bronze sample.  The Bronze sample includes 23 novae for which the fractional error in the {\it Gaia} parallax is $>$30\%, and so the distances have large error bars.  The left-side panel gives $D_{prior}$ versus $D_{Gaia}$, while the right-side panel gives $\varpi_{prior}$ versus $\varpi_{Gaia}$.  In an ideal world, the points would be tightly clustered around the $D_{prior}$=$D_{Gaia}$ diagonal line.  What we see is a huge scatter.  This confirms the conclusions from Figure 7.  However, in this figure, the error bars are sufficiently large that we cannot use the Bronze sample alone to convict the MMRD of failure.}
\end{figure*}

\begin{figure*}
	\includegraphics[width=\columnwidth]{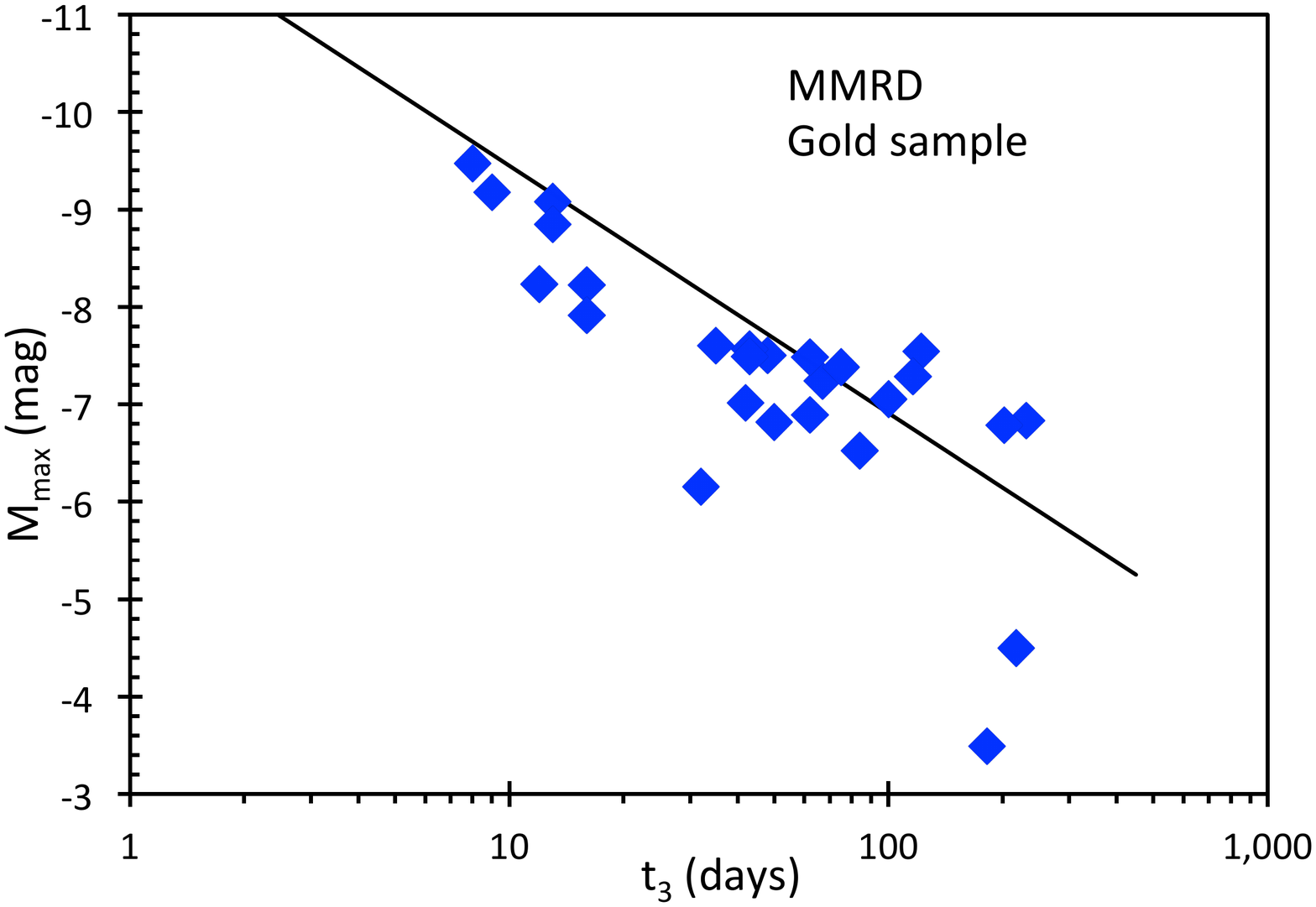}
	\includegraphics[width=\columnwidth]{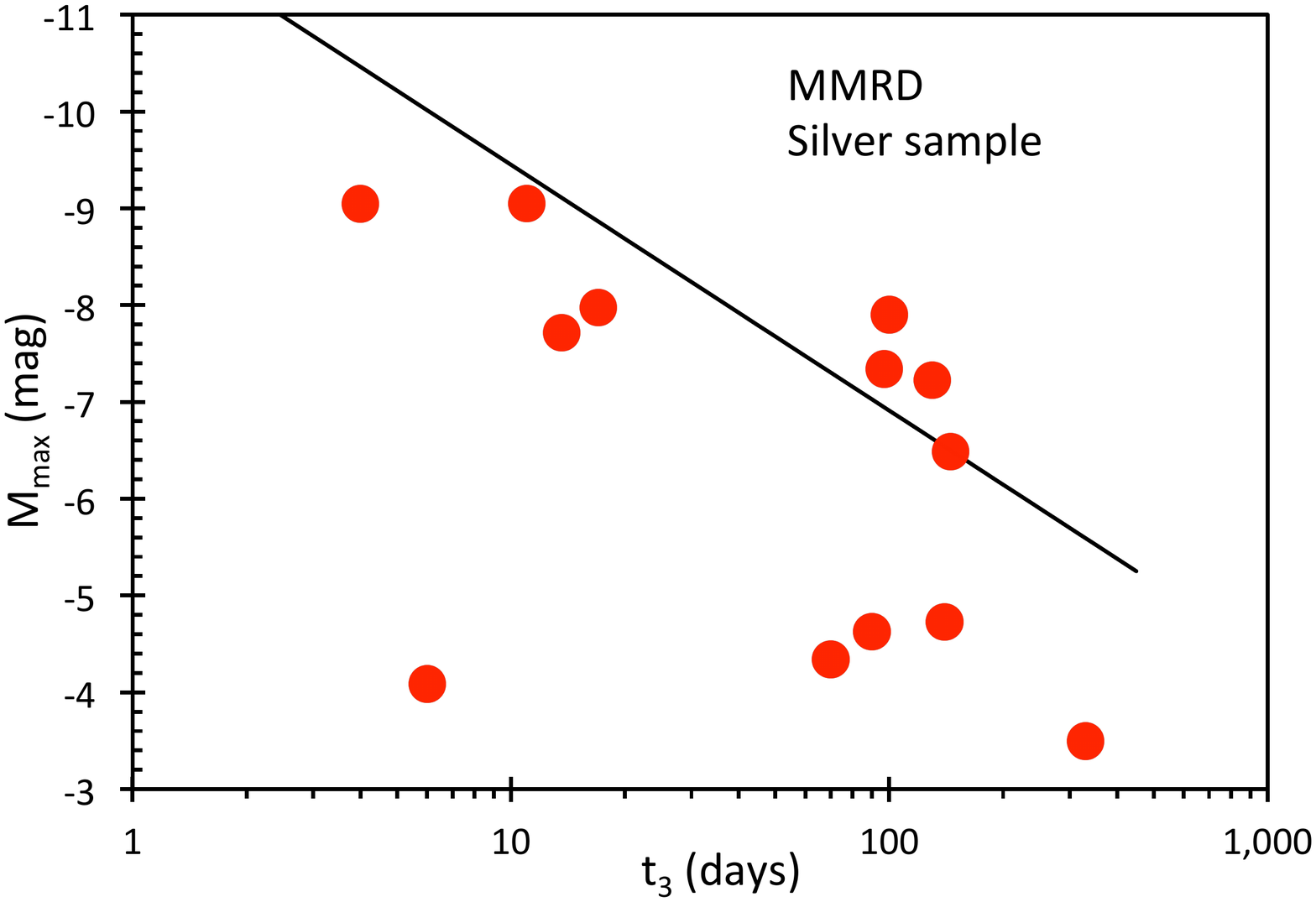}
    \caption{The MMRD for the Gold and Silver samples separately.  These two plots show the MMRD for the Gold sample (the panel on the left) and the Silver sample (the panel on the right).  Both of these plots can be directly compared to Figure 4a, with the same range and scale for both axes, and with the MMRD relation from Downes \& Duerbeck (solid black line).  We see that the Gold sample does display an apparent relation between $M_{max}$ and $t_3$, but there is huge scatter including several far outliers, and the best fit relation is substantially different from the pre-{\it Gaia} relation.  So if we willfully paid attention only to the Gold sample, then we could continue the old faith that we can pull out some utility from the MMRD.  But the Silver sample shows no relation at all between $M_{max}$ and $t_3$.  Rather, we see much of the diagram filled in, so the existence of the MMRD is denied.  The utter failure of the MMRD as shown in the Silver sample is also shown for novae in M31, for novae in M87, for galactic recurrent novae, and for grids of theoretical models of novae.  So we are faced with the dilemma of reconciling the two panels in this figure, where the Gold sample shows a real MMRD relation (albeit with so much scatter as to be essentially useless) while the Silver sample denies the existence of any MMRD.}
\end{figure*}

\begin{table*}
	\centering
	\caption{{\it Gaia} novae, parallaxes, distances, magnitudes, extinctions, and absolute magnitudes}
	\begin{tabular}{lllllllllllllll} 
		\hline
		Nova &
		Sample &
		Type &
		Year &
		LC & 
		$\varpi _{Gaia}$ (mas) &
		$D_{Gaia}$ (pc) &
		$V_{max}$ &
		$V_{15d}$ &
		$V_q$ &
		$A_V$ &
		$t_2$ &
		$t_3$ &
		$M_{max}$ &
		$M_q$
		\\
		\hline
CI Aql	&	Gold	&	RN	&	2000	&	P	&	0.327	$\pm$	0.050	&	3189	$_{-	315	}^{+	949	}$ &	9.0	&	10.2	&	16.1	&	2.6	&	25	&	32	&	-6.2	&	0.9	\\
V603 Aql	&	Gold	&	CN	&	1918	&	O	&	3.191	$\pm$	0.069	&	314	$_{-	7	}^{+	7	}$ &	-0.5	&	2.6	&	10.9	&	0.2	&	5	&	12	&	-8.2	&	3.2	\\
V1494 Aql	&	Gold	&	CN	&	1999	&	O	&	0.839	$\pm$	0.141	&	1239	$_{-	127	}^{+	422	}$ &	4.1	&	7.0	&	17.1	&	1.9	&	8	&	16	&	-8.2	&	4.8	\\
T Aur	&	Gold	&	CN	&	1891	&	D	&	1.141	$\pm$	0.051	&	880	$_{-	35	}^{+	46	}$ &	4.5	&	4.9	&	14.9	&	1.3	&	80	&	84	&	-6.5	&	3.9	\\
V705 Cas	&	Gold	&	CN	&	1993	&	D	&	0.480	$\pm$	0.087	&	2157	$_{-	228	}^{+	799	}$ &	5.7	&	6.6	&	16.4	&	1.3	&	33	&	67	&	-7.2	&	3.5	\\
V842 Cen	&	Gold	&	CN	&	1986	&	D	&	0.731	$\pm$	0.050	&	1379	$_{-	78	}^{+	120	}$ &	4.9	&	5.6	&	15.8	&	1.7	&	43	&	48	&	-7.5	&	3.4	\\
V476 Cyg	&	Gold	&	CN	&	1920	&	D	&	1.524	$\pm$	0.168	&	665	$_{-	53	}^{+	107	}$ &	1.9	&	4.8	&	16.2	&	0.7	&	6	&	16	&	-7.9	&	6.4	\\
V1330 Cyg	&	Gold	&	CN	&	1970	&	S	&	0.351	$\pm$	0.102	&	2883	$_{-	354	}^{+	1937	}$ &	9.9	&	10.1	&	17.5	&	2.1	&	161	&	217	&	-4.5	&	3.1	\\
V1974 Cyg	&	Gold	&	CN (Hi-$\Delta$m)	&	1992	&	P	&	0.617	$\pm$	0.069	&	1631	$_{-	131	}^{+	261	}$ &	4.3	&	5.9	&	$>$21	&	0.8	&	19	&	43	&	-7.6	&	$>$9.1	\\
HR Del	&	Gold	&	CN	&	1967	&	J	&	1.045	$\pm$	0.035	&	958	$_{-	29	}^{+	35	}$ &	3.6	&	3.8	&	12.1	&	0.5	&	167	&	231	&	-6.8	&	1.7	\\
DN Gem	&	Gold	&	CN	&	1912	&	P	&	0.729	$\pm$	0.081	&	1365	$_{-	108	}^{+	209	}$ &	3.6	&	5.5	&	15.6	&	0.5	&	16	&	35	&	-7.6	&	4.4	\\
DQ Her	&	Gold	&	CN	&	1934	&	D	&	1.997	$\pm$	0.024	&	501	$_{-	6	}^{+	6	}$ &	1.6	&	2.0	&	14.3	&	0.2	&	76	&	100	&	-7.1	&	5.6	\\
V446 Her	&	Gold	&	CN\&DN	&	1960	&	S	&	0.744	$\pm$	0.072	&	1361	$_{-	100	}^{+	185	}$ &	4.8	&	6.3	&	16.1	&	1.1	&	20	&	42	&	-7.0	&	4.3	\\
V533 Her	&	Gold	&	CN	&	1963	&	S	&	0.830	$\pm$	0.032	&	1202	$_{-	41	}^{+	52	}$ &	3.0	&	4.0	&	15.0	&	0.1	&	30	&	43	&	-7.5	&	4.5	\\
CP Lac	&	Gold	&	CN	&	1936	&	S	&	0.859	$\pm$	0.042	&	1170	$_{-	50	}^{+	67	}$ &	2.0	&	5.3	&	15.0	&	0.8	&	5	&	9	&	-9.2	&	3.8	\\
DK Lac	&	Gold	&	CN	&	1950	&	J	&	0.403	$\pm$	0.070	&	2517	$_{-	261	}^{+	788	}$ &	5.9	&	6.4	&	13.8	&	0.7	&	55	&	202	&	-6.8	&	1.1	\\
BT Mon	&	Gold	&	CN	&	1939	&	F	&	0.682	$\pm$	0.047	&	1477	$_{-	84	}^{+	128	}$ &	8.1	&	8.4	&	15.7	&	0.7	&	118	&	182	&	-3.5	&	4.1	\\
GK Per	&	Gold	&	CN\&DN	&	1901	&	O	&	2.263	$\pm$	0.043	&	442	$_{-	8	}^{+	9	}$ &	0.2	&	3.2	&	13.0	&	1.1	&	6	&	13	&	-9.1	&	3.7	\\
RR Pic	&	Gold	&	CN	&	1925	&	J	&	1.955	$\pm$	0.030	&	511	$_{-	8	}^{+	8	}$ &	1.0	&	1.4	&	12.2	&	0.0	&	73	&	122	&	-7.5	&	3.7	\\
CP Pup	&	Gold	&	CN (Hi-$\Delta$m)	&	1942	&	P	&	1.230	$\pm$	0.021	&	814	$_{-	14	}^{+	15	}$ &	0.7	&	3.9	&	$>$18.0	&	0.6	&	4	&	8	&	-9.5	&	$>$7.8	\\
T Pyx	&	Gold	&	RN (Hi-$\Delta$m)	&	2011	&	P	&	0.305	$\pm$	0.042	&	3185	$_{-	283	}^{+	607	}$ &	6.4	&	7.3	&	18.5	&	0.8	&	32	&	62	&	-6.9	&	5.2	\\
V732 Sgr	&	Gold	&	CN	&	1936	&	D	&	0.578	$\pm$	0.082	&	1795	$_{-	170	}^{+	458	}$ &	6.4	&	6.9	&	16.0	&	2.5	&	65	&	75	&	-7.4	&	2.2	\\
FH Ser	&	Gold	&	CN	&	1970	&	D	&	0.951	$\pm$	0.077	&	1060	$_{-	68	}^{+	112	}$ &	4.5	&	5.1	&	16.8	&	1.9	&	49	&	62	&	-7.5	&	4.8	\\
V382 Vel	&	Gold	&	CN	&	1999	&	S	&	0.560	$\pm$	0.055	&	1800	$_{-	133	}^{+	243	}$ &	2.8	&	5.8	&	16.6	&	0.4	&	6	&	13	&	-8.8	&	5.0	\\
NQ Vul	&	Gold	&	CN	&	1976	&	D	&	0.946	$\pm$	0.100	&	1080	$_{-	85	}^{+	169	}$ &	6.2	&	7.6	&	17.2	&	2.9	&	21	&	50	&	-6.8	&	4.2	\\
PW Vul	&	Gold	&	CN	&	1984	&	J	&	0.426	$\pm$	0.100	&	2420	$_{-	277	}^{+	1337	}$ &	6.4	&	7.1	&	16.9	&	1.8	&	44	&	116	&	-7.3	&	3.2	\\
V368 Aql	&	Silver	&	CN	&	1936	&	S	&	0.371	$\pm$	0.052	&	2722	$_{-	253	}^{+	612	}$ &	5.0	&	7.7	&	16.6	&	0.8	&	5	&	17	&	-8.0	&	3.6	\\
Z Cam	&	Silver	&	CN\&DN	&	77	&	...	&	4.437	$\pm$	0.040	&	225	$_{-	1	}^{+	4	}$ &	0$\pm$3	&	...	&	13.6	&	0.1	&	...	&	...	&	-6.8	&	6.8	\\
AT Cnc	&	Silver	&	CN\&DN	&	1645	&	...	&	2.201	$\pm$	0.047	&	454	$_{-	9	}^{+	10	}$ &	0$\pm$3	&	...	&	13.9	&	0.0	&	...	&	...	&	-8.3	&	5.6	\\
BC Cas	&	Silver	&	CN	&	1929	&	...	&	0.490	$\pm$	0.071	&	2114	$_{-	203	}^{+	557	}$ &	10.7	&	...	&	17.0	&	3.7	&	...	&	90	&	-4.6	&	1.7	\\
AR Cir	&	Silver	&	CN (Symb?)	&	1906	&	...	&	0.578	$\pm$	0.120	&	1886	$_{-	186	}^{+	1478	}$ &	10.3	&	...	&	18.3	&	2.4	&	...	&	330	&	-3.5	&	4.5	\\
Q Cyg	&	Silver	&	CN	&	1876	&	...	&	0.729	$\pm$	0.024	&	1372	$_{-	42	}^{+	51	}$ &	3.0	&	...	&	15.0	&	1.4	&	...	&	11	&	-9.1	&	2.9	\\
KT Eri	&	Silver	&	CN	&	2009	&	P	&	0.204	$\pm$	0.038	&	3744	$_{-	328	}^{+	591	}$ &	5.4	&	8.4	&	15.0	&	0.2	&	7	&	14	&	-7.7	&	1.9	\\
HR Lyr	&	Silver	&	CN	&	1919	&	S	&	0.182	$\pm$	0.034	&	4797	$_{-	470	}^{+	1015	}$ &	6.5	&	7.5	&	15.5	&	0.4	&	47	&	97	&	-7.3	&	1.7	\\
V841 Oph	&	Silver	&	CN	&	1848	&	J	&	1.215	$\pm$	0.027	&	823	$_{-	17	}^{+	20	}$ &	4.3	&	4.9	&	13.4	&	1.2	&	50	&	145	&	-6.5	&	2.6	\\
V392 Per	&	Silver	&	CN\&DN	&	2018	&	P	&	0.257	$\pm$	0.052	&	4161	$_{-	440	}^{+	2345	}$ &	5.6	&	10.2	&	17.4	&	1.6	&	2	&	4	&	-9.0	&	2.8	\\
HZ Pup	&	Silver	&	CN	&	1963	&	J	&	0.408	$\pm$	0.067	&	2560	$_{-	260	}^{+	851	}$ &	7.7	&	9.3	&	16.9	&	0.0	&	60	&	70	&	-4.3	&	4.9	\\
V1017 Sgr	&	Silver	&	CN\&DN	&	1919	&	S	&	0.789	$\pm$	0.044	&	1269	$_{-	60	}^{+	84	}$ &	4.5$\pm$2.0	&	...	&	13.5	&	1.2	&	...	&	130	&	-7.2	&	1.8	\\
V1016 Sgr	&	Silver	&	CN	&	1899	&	...	&	0.377	$\pm$	0.032	&	2664	$_{-	175	}^{+	291	}$ &	8.5	&	8.6	&	14.0	&	1.1	&	64	&	140	&	-4.7	&	0.8	\\
V721 Sco	&	Silver	&	CN	&	1950	&	...	&	0.641	$\pm$	0.051	&	1574	$_{-	100	}^{+	165	}$ &	8.0	&	13.5	&	16.0	&	1.1	&	...	&	6	&	-4.1	&	3.9	\\
CT Ser	&	Silver	&	CN	&	1948	&	J	&	0.230	$\pm$	0.063	&	2774	$_{-	268	}^{+	495	}$ &	$\sim$5	&	...	&	16.6	&	0.7	&	...	&	100	&	-7.9	&	3.7	\\
OS And	&	Bronze	&	CN	&	1986	&	D	&	0.138	$\pm$	0.138	&	3298	$_{-	524	}^{+	1670	}$ &	6.5	&	8.8	&	17.5	&	0.3	&	11	&	23	&	-6.4	&	4.6	\\
V356 Aql	&	Bronze	&	CN	&	1936	&	J	&	0.476	$\pm$	0.209	&	2427	$_{-	285	}^{+	3672	}$ &	7.0	&	7.2	&	18.3	&	2.0	&	127	&	140	&	-6.9	&	4.4	\\
V1229 Aql	&	Bronze	&	CN	&	1970	&	P	&	0.650	$\pm$	0.665	&	2786	$_{-	712	}^{+	4213	}$ &	6.6	&	8.3	&	18.1	&	1.6	&	18	&	32	&	-7.2	&	4.3	\\
V1370 Aql	&	Bronze	&	CN	&	1982	&	D	&	0.339	$\pm$	0.189	&	2928	$_{-	450	}^{+	3198	}$ &	7.7	&	9.7	&	18.0	&	1.1	&	15	&	28	&	-5.7	&	4.6	\\
QZ Aur	&	Bronze	&	CN	&	1964	&	S	&	0.349	$\pm$	0.108	&	3200	$_{-	330	}^{+	4030	}$ &	5	&	...	&	17.0	&	1.7	&	...	&	26	&	-9.2	&	2.7	\\
V723 Cas	&	Bronze	&	CN (Hi-$\Delta$m)	&	1995	&	J	&	0.131	$\pm$	0.047	&	5628	$_{-	710	}^{+	1912	}$ &	7.1	&	7.2	&	18.8	&	1.4	&	263	&	299	&	-8.0	&	3.6	\\
V868 Cen	&	Bronze	&	CN	&	1991	&	J	&	1.145	$\pm$	0.881	&	14800	$_{-	4900	}^{+	20600	}$ &	8.7	&	9.7	&	19.9	&	5.3	&	31	&	82	&	-12.5	&	-1.3	\\
V888 Cen	&	Bronze	&	CN	&	1995	&	O	&	0.341	$\pm$	0.117	&	3376	$_{-	307	}^{+	5457	}$ &	8.0	&	8.8	&	15.2	&	1.1	&	38	&	90	&	-5.7	&	1.5	\\
BY Cir	&	Bronze	&	CN	&	1995	&	P	&	0.302	$\pm$	0.170	&	3804	$_{-	560	}^{+	5434	}$ &	7.4	&	8.3	&	17.9	&	0.4	&	35	&	124	&	-5.9	&	4.6	\\
V2275 Cyg	&	Bronze	&	CN	&	2001	&	S	&	0.217	$\pm$	0.213	&	10800	$_{-	2973	}^{+	16500	}$ &	6.9	&	10.2	&	18.4	&	3.1	&	3	&	8	&	-11.4	&	0.1	\\
V2491 Cyg	&	Bronze	&	CN	&	2008	&	C	&	0.043	$\pm$	0.114	&	6517	$_{-	1238	}^{+	4911	}$ &	7.5	&	10.4	&	20.0	&	0.7	&	4	&	16	&	-7.3	&	5.2	\\
V339 Del	&	Bronze	&	CN	&	2013	&	P	&	0.381	$\pm$	0.361	&	2130	$_{-	400	}^{+	2250	}$ &	4.5	&	6.8	&	17.5	&	0.6	&	12	&	22	&	-7.7	&	5.3	\\
V838 Her	&	Bronze	&	CN (RN?)	&	1991	&	P	&	0.464	$\pm$	0.711	&	2530	$_{-	636	}^{+	3434	}$ &	5.3	&	12.0	&	19.1	&	1.6	&	1	&	4	&	-8.3	&	5.5	\\
GQ Mus	&	Bronze	&	CN (Hi-$\Delta$m)	&	1983	&	P	&	0.470	$\pm$	0.219	&	2480	$_{-	300	}^{+	3780	}$ &	7.2	&	8.8	&	21	&	1.395	&	35	&	45	&	-6.2	&	7.6	\\
IM Nor	&	Bronze	&	RN	&	2002	&	P	&	0.995	$\pm$	0.415	&	1205	$_{-	119	}^{+	2116	}$ &	8.5	&	9.1	&	18.3	&	2.5	&	50	&	80	&	-4.4	&	5.4	\\
V849 Oph	&	Bronze	&	CN	&	1919	&	F	&	0.153	$\pm$	0.200	&	2643	$_{-	445	}^{+	1531	}$ &	7.6	&	7.8	&	18.8	&	0.2	&	140	&	270	&	-4.7	&	6.5	\\
V2487 Oph	&	Bronze	&	RN	&	1998	&	P	&	0.113	$\pm$	0.099	&	4669	$_{-	748	}^{+	2558	}$ &	9.5	&	13.0	&	17.7	&	1.6	&	6	&	8	&	-5.4	&	2.8	\\
V351 Pup	&	Bronze	&	CN	&	1991	&	P	&	0.086	$\pm$	0.511	&	15900	$_{-	4900	}^{+	21100	}$ &	6.4	&	8.8	&	19.6	&	2.2	&	9	&	26	&	-11.8	&	1.4	\\
U Sco	&	Bronze	&	RN	&	2010	&	P	&	-0.352	$\pm$	0.215	&	19600	$_{-	5300	}^{+	21000	}$ &	7.5	&	15.0	&	17.6	&	0.6	&	1	&	3	&	-9.6	&	0.5	\\
V992 Sco	&	Bronze	&	CN	&	1992	&	D	&	0.397	$\pm$	0.150	&	3030	$_{-	236	}^{+	5960	}$ &	7.7	&	8.0	&	17.2	&	4.0	&	100	&	120	&	-8.7	&	0.8	\\
RW UMi	&	Bronze	&	CN (Hi-$\Delta$m)	&	1956	&	...	&	0.472	$\pm$	0.193	&	1510	$_{-	199	}^{+	564	}$ &	6	&	...	&	$>$21	&	0.09	&	...	&	140	&	-5.0	&	$>$10.0	\\
QU Vul	&	Bronze	&	CN	&	1984	&	P	&	0.739	$\pm$	0.366	&	1786	$_{-	196	}^{+	3495	}$ &	5.3	&	6.8	&	17.9	&	1.7	&	20	&	36	&	-7.7	&	4.9	\\
QV Vul	&	Bronze	&	CN	&	1987	&	D	&	0.125	$\pm$	0.213	&	3619	$_{-	694	}^{+	3058	}$ &	7.1	&	7.9	&	18.0	&	1.2	&	37	&	47	&	-6.9	&	4.0	\\
		\hline
	\end{tabular}
\end{table*}

\newpage
\begin{table}
	\centering
	\caption{HST FGS parallaxes to four classical novae}
	\begin{tabular}{lll} 
		\hline
		Nova & $\varpi _{HST}$ (mas) & $\varpi _{Gaia}$ (mas)\\
		\hline
V603 Aql	&	4.011	$\pm$	0.137	&	3.191	$\pm$	0.069	\\
DQ Her	&	2.594	$\pm$	0.207	&	1.997	$\pm$	0.024	\\
GK Per	&	2.097	$\pm$	0.116	&	2.263	$\pm$	0.043	\\
RR Pic	&	1.920	$\pm$	0.182	&	1.955	$\pm$	0.030	\\
		\hline
	\end{tabular}
\end{table}

\newpage

\begin{table}
	\centering
	\caption{Published Expansion Parallaxes Compared to {\it Gaia}}
	\begin{tabular}{lll} 
		\hline
		Nova & $D_{Gaia}$ (pc) & $D_{prior}$ (pc)\\
		\hline
V603 Aql	&	314	$_{-	7	}^{+	7	}$ &	330, 400	\\
V1494 Aql	&	1239	$_{-	127	}^{+	422	}$ &	1200	\\
T Aur	&	880	$_{-	35	}^{+	46	}$ &	960$\pm$220, 1500	\\
V705 Cas	&	2157	$_{-	228	}^{+	799	}$ &	2500	\\
V842 Cen	&	1379	$_{-	78	}^{+	120	}$ &	420, 1200, 1140, 1150, 1300$\pm$500, 1325	\\
V476 Cyg	&	665	$_{-	53	}^{+	107	}$ &	1620$\pm$120, 1800	\\
V1974 Cyg	&	1631	$_{-	131	}^{+	261	}$ &	310, 1320, 1800$\pm$100, 1940, 2000, 2115,  \\
  &  &~~~~~$\sim$2600 	\\
HR Del	&	958	$_{-	29	}^{+	35	}$ &	 505, 750, 760$\pm$130, 825, 900, 940$\pm$155,  \\
  &  &~~~~~ 1010	\\
DQ Her	&	501	$_{-	6	}^{+	6	}$ &	300, 400$\pm$60, 480$\pm$50, 560$\pm$20	\\
V446 Her	&	1361	$_{-	100	}^{+	185	}$ &	1350	\\
V533 Her	&	1202	$_{-	41	}^{+	52	}$ &	560$\pm$70, 850$\pm$150, 1250$\pm$30, 1300, 1320	\\
CP Lac	&	1170	$_{-	50	}^{+	67	}$ &	1350	\\
DK Lac	&	2517	$_{-	261	}^{+	788	}$ &	3900$\pm$500	\\
BT Mon	&	1477	$_{-	84	}^{+	128	}$ &	1800$\pm$300	\\
GK Per	&	442	$_{-	8	}^{+	9	}$ &	455$\pm$30, 500	\\
RR Pic	&	511	$_{-	8	}^{+	8	}$ &	500, 580, 600$\pm$60	\\
CP Pup	&	814	$_{-	14	}^{+	15	}$ &	850, 900, 1120, 1140, 1500, 1600, 1700,  \\
  &  &~~~~~ 1800$\pm$400	\\
FH Ser	&	1060	$_{-	68	}^{+	112	}$ &	700, 850$\pm$50, 870$\pm$90, 920$\pm$130, 950$\pm$50	\\
V382 Vel	&	1800	$_{-	133	}^{+	243	}$ &	800	\\
NQ Vul	&	1080	$_{-	85	}^{+	169	}$ &	910, 1160$\pm$210, 1200, 1280, 1600$\pm$800,  \\
  &  &~~~~~ 1700	\\
PW Vul	&	2420	$_{-	277	}^{+	1337	}$ &	1200, 1300, 1500-3000, 1600, 1635, 1750,  \\
  &  &~~~~~ 1800$\pm$50, 1880	\\
CT Ser	&	2774	$_{-	268	}^{+	495	}$ &	1430	\\
		\hline
	\end{tabular}
\end{table}

\begin{table}
	\centering
	\caption{Collected results for quality of prior nova distances}
	\begin{tabular}{llllll} 
		\hline
		Prior Method & Nova Sample & $N_{nova}$  &   $\sigma_{\Delta \mu}$      &      $\sigma_{\chi}$      &        $\langle \Delta \mu \rangle$      \\
		\hline
{\it HST} FGS parallaxes	&	Harrison et al. (2013)	&	4	&	0.37	&	3.03	&	-0.21	\\
Ground-based parallaxes	&	CVs of Thorstensen	&	26	&	0.54	&	1.06	&	-0.37	\\
Expansion parallaxes	&	Slavin (1997)	&	9	&	1.04	&	3.60	&	-0.10	\\
Expansion parallaxes	&	Table 3	&	75	&	0.95	&	$^a$	&	-0.06	\\
Blackbody distance to secondary	&	V1017 Sgr (Salazar et al. 2016)	&	1	&	...	&	$\chi$=-0.14	&	-0.05	\\
Reddening-distance relation	&	\"{O}zd\"{o}nmez et al. (2016)	&	15	&	1.32	&	4.25	&	-0.05	\\
Reddening-distance relation	&	\"{O}zd\"{o}nmez et al. (2018)	&	5	&	1.01	&	1.31	&	0.12	\\
RDR and parallaxes	&	\"{O}zd\"{o}nmez et al. (2018)	&	28	&	1.15	&	2.45	&	-0.05	\\
MMRD for $t_3$	&	Gold+Silver	&	39	&	1.31	&	1.03	&	0.73	\\
MMRD for $t_3$	&	Gold	&	26	&	0.82	&	0.83	&	0.43	\\
MMRD for $t_3$	&	Silver	&	13	&	1.85	&	1.32	&	1.32	\\
MMRD for $t_3$	&	Bronze	&	23	&	2.20	&	1.68	&	0.44	\\
$M_{15d}$ = -5.23$\pm$0.39 mag	&	Gold+Silver	&	34	&	1.53	&	2.26	&	0.04	\\
$M_{15d}$ = -5.23$\pm$0.39 mag	&	Gold+Silver ($t_2$$>$10 days)	&	23	&	1.08	&	2.21	&	-0.28	\\
$M_{max}$ = -7.0$\pm$1.4 mag	&	Gold+Silver	&	37	&	1.61	&	$^b$	&	0.00	\\
Nova light echo	&	T Pyx (Sokoloski et al. 2013)	&	1	&	...	&	$\chi$=2.05	&	0.89	\\
Combining multiple methods	&	Schaefer papers	&	5	&	0.93	&	0.86	&	0.82	\\
Combining multiple methods	&	Duerbeck (1981)	&	17	&	1.05	&	4.38	&	-0.43	\\
Combining multiple methods	&	Patterson (1984)	&	9	&	0.42	&	$^a$	&	0.04	\\
Combining multiple methods	&	Patterson et al. (2013)	&	13	&	0.44	&	$^a$	&	0.17	\\
Theory models	&	Hachisu \& Kato papers	&	16	&	0.76	&	1.66	&	0.16	\\
		\hline
	\end{tabular}
$^a$Formal uncertainties are not available for most novae in the set. \\
$^b$The $\pm$1.4 value was chosen so that $\sigma_{\chi}$=1.00.\\
\end{table}

\bsp	
\label{lastpage}
\end{document}